\documentclass[a4,latin,onecolumn]{paper}
\usepackage[utf8]{inputenc}
\usepackage[margin=2.5cm]{geometry}
\usepackage{authblk}
\usepackage{hyperref}
\usepackage{array}
\usepackage{graphicx}
\usepackage[T1]{fontenc}
\usepackage{subcaption}
\usepackage{amsmath}   
\usepackage{amssymb}   
\usepackage{amsfonts}  
\usepackage{xcolor}
\usepackage{graphicx}
\usepackage{algorithm2e}
\usepackage{tcolorbox}
\usepackage{cite}
\usepackage{placeins}
\usepackage{setspace}

\begin{document}
\title{\centering{Information Entropy Based Crystal Structure Prediction of Chemically Disordered Alloys via Graph Convolutional Neural Networks}}
\author{\normalsize {\textbf {Suman Chabri$^1$ and Gautam Anand$^2$}} \thanks{Email: gautamanand.mst@itbhu.ac.in}}
\affil[1]{Department of Metallurgy and Materials Engineering, Indian Institute of Engineering Science and Technology, Shibpur, Howrah, WB 711103, India}
\affil[2]{School of Materials Science and Technology, Indian Institute of Technology (BHU), Varanasi 221005, India}

\date{}
\maketitle

\abstract
The phase prediction of chemically disordered alloys poses a significant computational challenge due to the combinatorial complexity of such materials. The high-throughput compositional exploration of chemically disordered alloys, including high-entropy alloys, requires an approach to efficiently explore the potential energy landscape of such complex materials. Additionally, a metric to quantify the potential energy landscape explored for phase prediction of the compositions needs to be defined.  We propose an information-theoretic approach to phase prediction in chemically disordered alloys in the present work. We demonstrate the applicability of alchemical Monte Carlo sampling using an efficient Graph Convolutional Neural Network-Based machine learning model. We additionally demonstrate the applicability and limitations of the Bond Disproportion Vector (BDV) as a low-computational-cost descriptor and benchmark it against the state-of-the-art Smooth Overlap of Atomic Positions (SOAP) descriptor. We show the applicability of an information entropy-based metric for the phase prediction of binary (CoNi, MoW, FeNi and TaW), ternary (CoCrNi, CrFeNi), quaternary (CoCrFeNi) and quinary ($\mathrm{Al_x(CoCrFeNi)_{1-x}}$) alloys. Information entropy-based phase prediction can be applicable in challenging cases where conventional approaches are not feasible. 
  
\doublespacing
\section{Introduction}
High-entropy alloys (HEA) and other chemically complex, multicomponent alloys offer a vast compositional design space but pose a fundamental challenge for predicting crystal structures. The high degree of chemical disorder in these systems leads to enormous configurational complexity, making conventional first-principles methods too costly for the high-throughput exploration of candidate compositions and structures \cite{guo2025high,woodgate2022compositional,qureshi2025predictive,bista2025fast}. The thermodynamic CALPHAD-based approaches have limited applicability for the unknown candidate compositions and structures \cite{lacour2020effect,reynolds2020comparing}.  At the same time, phase stability and crystal structure selection are central to determining the structural \cite{almisned2025comprehensive} and functional properties \cite{yin2023valence} of these materials and hence, reliable prediction tools are crucial for accelerating alloy discovery and optimisation.\\
Recent advances in machine learning for atomistic modelling, particularly graph-based neural architectures, provide a promising route to address this challenge by directly learning structure-energy relationships from data on disordered configurations. Graph neural networks (GNN) have been successfully applied to energetic and structural predictions across composition spaces in refractory solid-solution alloys  \cite{pasini2025transferable}, nested crystal graph models for chemically complex systems  \cite{wang2025nested}, and phase-specific enthalpy predictions in intermetallics \cite{zhang2025prediction}, underscoring their versatility for complex alloy chemistry. The GNN has been applied for powering a large language model-driven multi-agent artificial intelligence framework for HEA property prediction \cite{ghafarollahi2025rapid},  designing bimettalic systems \cite{gu2025high}, prediction of hydrogen storage property of metal-organic frameworks \cite{lu2022hydrogen},  grain boundary structure search \cite{zhang2023graph}, prediction of properties of polycrystalline materials \cite{dai2021graph}, composition-based property prediction \cite{fung2021benchmarking}, and properties prediction in general \cite{louis2020graph,shi2024review,merchant2023scaling,chen2023md}. Graph convolutional neural networks (GCNN) treat atomistic structures as graphs, with atoms as nodes and bonds or near-neighbour relations as edges, enabling efficient message passing on irregular, non-grid-like data and capturing local chemical environments in a flexible way that circumvents the limitations of conventional grid-based convolutional networks for atomistic inputs. \\
Building on these developments, the present work introduces an information-entropy-based framework that combines alchemical Monte Carlo-style atomistic sampling, GCNN-based energy prediction, and a new Bond Disproportion Vector (BDV) descriptor, benchmarked against the Smooth Overlap of Atomic Positions (SOAP) descriptor \cite{bartok2013representing}, to enable data-driven crystal structure prediction in chemically disordered alloys. The BDV encodes deviations of first-shell bond statistics from a fully random alloy as a compact vector, providing a low-dimensional yet physically interpretable representation tailored to chemically disordered environments. These descriptors are featurised using a Term Frequency-Inverse Document Frequency (TF-IDF) weighting scheme adopted from information retrieval to emphasise rare but energetically important coordination motifs in large atomistic datasets. Coupled with an alchemical Monte Carlo sampling protocol based on the Genetic Algorithm-based Atomistic Sampling Protocol (GAASP) for efficient exploration of thermodynamically relevant configurations, and an information-entropy metric grounded in Kullback-Leibler divergence is used to compare sampled energy distributions between candidate structures and equilibrium-like references; this framework aims to achieve robust, entropy-informed phase prediction in high-entropy and other chemically disordered alloys.

\section{Method}
\subsection{Descriptor for atomic environment} \label{descriptor}
The atomic descriptor captures an atom's local environment and can be used to predict atomic attributes using machine learning algorithms. The design of the descriptor plays a crucial role in the ML model's predictive accuracy. In the present investigation, the aim is to develop an ML algorithm for predicting the potential energy of an atomistic configuration in chemically disordered alloys. In such cases, the local atomic configuration shows significant statistical variation; hence, the developed descriptor should account for this variation and its effect on the potential energy of an atom in the atomistic configuration. \\ 
The Smooth Overlap of Atomic Positions (SOAP) represents the atomic environment as a smooth density field,
\begin{equation}
\mathrm{
	\centering
	\rho_i\left(r\right) = \sum_j{f_{cut}\left(r_{ij} \right) \exp \left(\frac{-(r - r_{ij})^2}{2\sigma^2} \right)}
	}
\end{equation}
where, $\mathrm{r_{ij}}$ is the distance between atom $\mathrm{i}$ and neighbour $\mathrm{j}$ atom, $\mathrm{\sigma}$ is the Gaussian width, while $\mathrm{f_{cut} (r_{ij})}$ is the cut-off function whose value goes to zero at $\mathrm{r = r_{cut}}$ (\emph{i.e.}, the cut-off radius). The density field can be decomposed into orthogonal radial and angular components (basis functions, in the same spirit as diffraction patterns are expanded in the reciprocal space basis vectors or vibrational modes are expressed as a superposition of normal modes),
\begin{equation}
\mathrm{
	\centering
	\rho(r) \approx \sum_{n,l,m}{c_{nlm}} g_{n}(r)Y_{lm}(\hat{r})
	}
\end{equation} 
The radial functions provide information about the variation of the atomic density with distance, \emph{i.e.}, the positions of the coordination shells. The spherical harmonics describe the directionality of the density, \emph{i.e.}, different bonding geometries or atomic-packing motifs can be distinguished. The expansion coefficients ($\mathrm{c_{nlm}}$) are then converted into the power spectrum ($\mathrm{p_{nn'l}}$) as,
\begin{equation}
\mathrm{
	\centering
	p_{nn'l} = \sum_{m=-l}^{l} {c_{nlm }c_{nn'l}^{*}}
	}
\end{equation}
The $\mathrm{*}$ in the above expression represents the complex conjugation. The power spectrum provides radial (coordination shell) and angular (bond geometry) information, which is rotationally invariant. Such a combined representation enables the expression of the atomic environment as a weighted sum of physically interpretable structural signatures. providing a compact and symmetry-preserving descriptor that systematically encodes the bond lengths as well as the bond angles. \\
In addition to the above, to address the challenge of predicting the atom's potential energy in a chemically disordered environment, we propose a descriptor based on the bond count in the first coordination shell. We introduce \emph{bond disproportion} as a quantitative measure of the excess or deficiency of a given bond type relative to its expected occurrence in a completely random alloy, where all bonds have equal probability of formation. Consider an alloy composed of species $\mathrm{\{A, B, C,\dots\}}$ and let the first coordination shell contain all possible nearest-neighbour bonds. The \emph{Bond Disproportion Vector} (BDV) contains one entry for each distinct bond type. Let $\mathrm{N_{ij}}$ denote the observed number of $\mathrm{i}$--$\mathrm{j}$ bonds in the simulated or experimental configuration, and let $\mathrm{N_{ij}}^{\mathrm{rand}}$ be the statistically expected number of $\mathrm{i}$--$\mathrm{j}$ bonds in a fully disordered state. The disproportion associated with bond type $\mathrm{i}$--$\mathrm{j}$ is defined as
\begin{equation}
\mathrm{
D_{ij} = \frac{N_{ij} - N_{ij}^{\mathrm{rand}}}{N_{ij}^{\mathrm{rand}}}.
}
\label{eq:bdv}
\end{equation}

The BDV is then written as
\begin{equation}
\mathrm{
\mathbf{D} = \left( D_{AA},\, D_{AB},\, D_{AC},\, \dots \right)
}
\end{equation}
A positive $\mathrm{D_{ij}}$ indicates enrichment of the $\mathrm{i}$--$\mathrm{j}$ bond relative to the random state, whereas a negative value indicates depletion. The BDV encodes only the statistical disproportion of bond counts and does not include geometrical information such as bond lengths or bond angles. The aim of employing the BDV in light of the state-of-the-art SOAP descriptor is to propose and benchmark a computationally efficient descriptor for chemically disordered alloys. Figure 1 in the supplementary information (SI) shows that the descriptor vector size increases for SOAP and BDV with increasing alloy element count. It is clear that as the number of alloy elements increases, the SOAP descriptor vector size increases significantly more than in BDV. The increased size of the descriptor vector for an atom is a crucial concern in high-throughput computation.
\subsection{Featurisation of the atomic descriptors} \label{feature}	
The atomic descriptors store the information about the atomic coordination. However, it is important to featurise the atomic descriptor vectors. The featurisation strategy should consider the machine learning requirements. In the present investigation, Term Frequency-Inverse Document Frequency (TF-IDF) weighting is employed in Natural Language Processing and information retrieval to measure the importance of a word within a corpus \cite{wang2024research}. We employ TF-IDF weighting for the featurization of the atomic descriptor (Fig. \ref{subfig:schmatic-method}), as TF-IDF is useful when the feature importance depends on the variety, which is the case for chemically disordered alloys. The TF-IDF is employed for large datasets where common patterns need to be downweighted, while critical components with significant effect need to be highlighted. In the case of atomistic systems, the atomic species and their corresponding coordination environments, which give rise to the energy variations, should be learned to predict the potential energy of the atomistic system. The TF-IDF contains two terms, TF (Term Frequency) and IDF (Inverse Document Frequency), which may be represented as,
\begin{equation}
\mathrm{
	TF = \frac{n_W}{t_W} \: \: \& \: \:
	IDF = log\left( \frac{t_D}{t_{DW}} + 1 \right)
}
\end{equation}
where, $\mathrm{n_W}$, $\mathrm{t_W}$, $\mathrm{t_D}$, $\mathrm{t_{DW}}$ are number of times a word $\mathrm{W}$ appears in the document, the total number of words in the document, total number of documents, and number of documents containing the term, respectively. In our work, we replace the words with the value in the atomic descriptor vector (\emph{i.e.}, bond disproportion in BDV or value of amplitude in the power spectrum in case of SOAP). The TF-IDF is then calculated as $\mathrm{TF \cdot IDF}$.  

\subsection{Graph Convolutional Neural Network (GCN) for Potential Energy Prediction}
\begin{figure}[h]
	\centering
	\begin{subfigure}{0.5\textwidth}
		\centering
		\includegraphics[width=\textwidth]{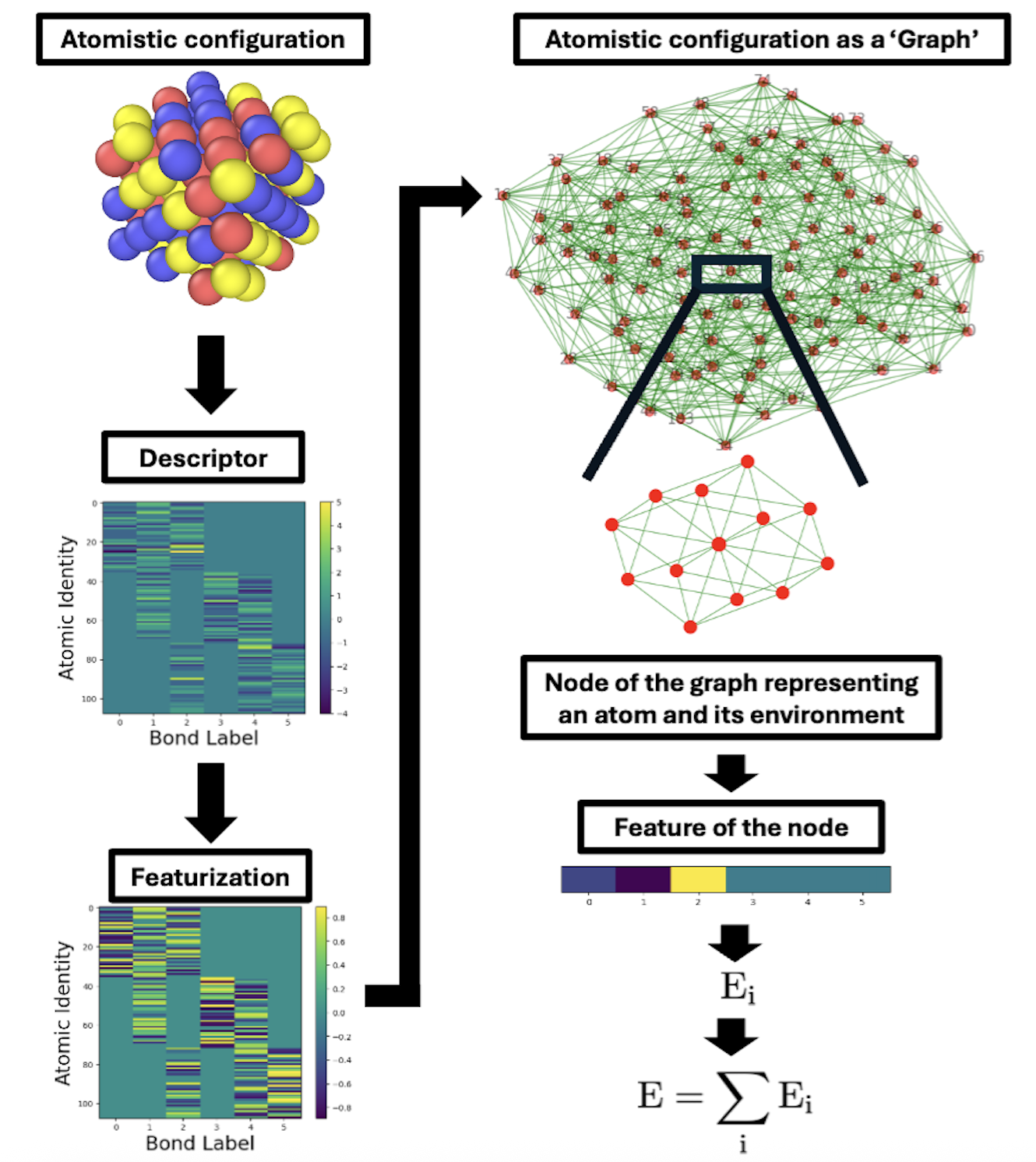}
		\caption{}
		\label{subfig:schmatic-method}
	\end{subfigure}
	\begin{subfigure}{0.47\textwidth}
		\centering
		\includegraphics[width=\textwidth]{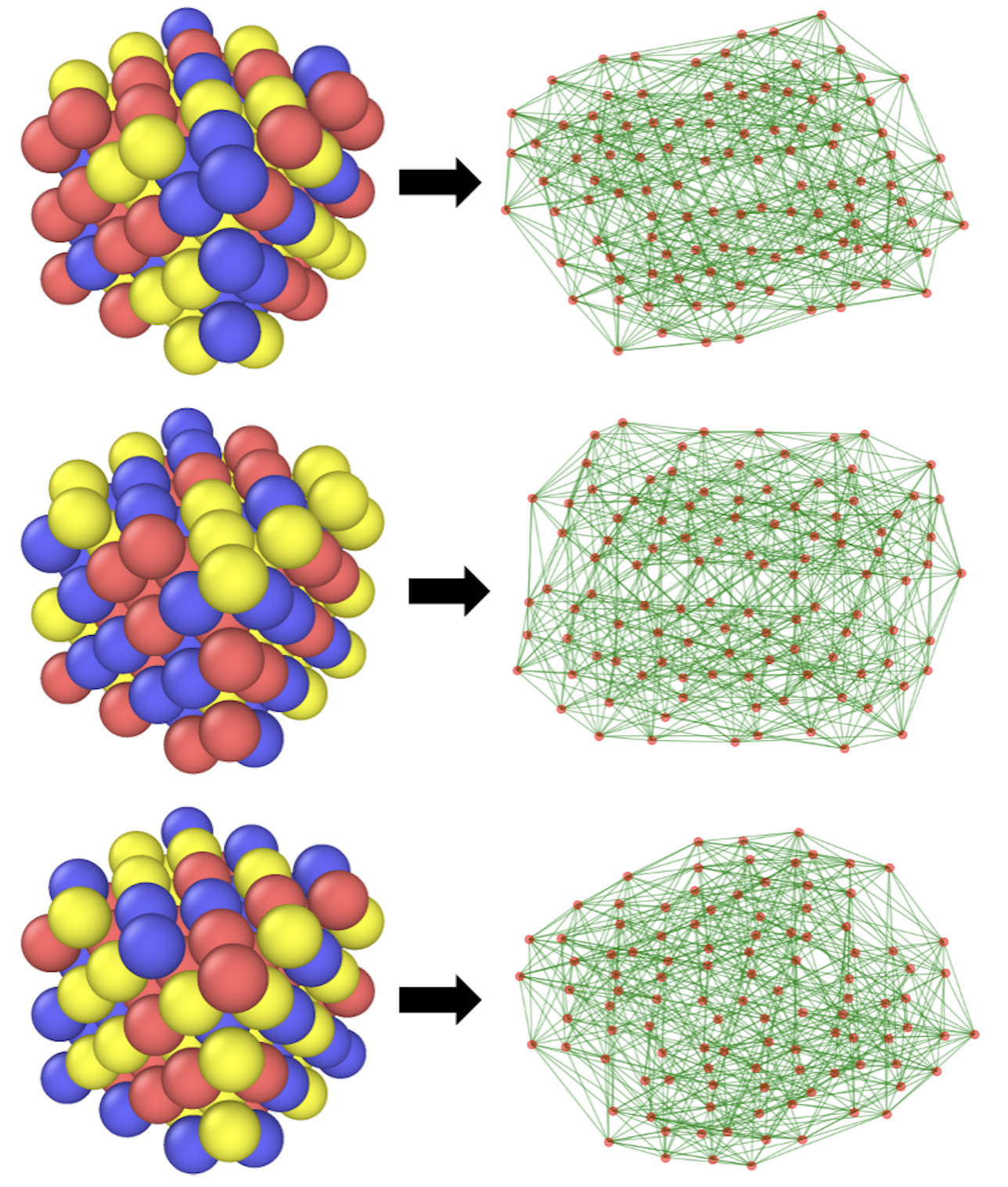}
		\caption{}
		\label{subfig:multiple}
	\end{subfigure}
\caption{(a) Schematic of the descriptor generation and featurisation of the atomistic configuration to generate a graph and (b) graphs for three different configurations of CoCrNi alloy showing the atomistic data leading to the unstructured graphs.}
\end{figure}

Predicting the energy of the atomistic structure poses a significant challenge, as each HEA atomistic configuration is chemically disordered. A change in the atomic environment can alter an atom's potential energy. The energy of each atom needs to be learned, which is further added to deduce the potential energy of the configuration. In machine learning, the potential energy of the atomistic structure poses few challenges: the input vector representing the descriptor is multidimensional, the scalar value needs to be learned, and the feature vectors representing the data do not exhibit a regular grid-like structure. The conventional neural network architecture, such as Artificial Neural Network (ANN), struggles with the multidimensional features. Hence, the Convolutional Neural Network (CNN) architecture is better suited to the scenario involving a multidimensional input feature vector and a scalar output value.  The conventional CNN, though useful for high-dimensional input-to-scalar-output scenarios, is better suited to regular-grid-like data, as is generally the case for images.  However, the atomistic configuration as a graph is inherently unstructured (Fig. \ref{subfig:multiple}). Hence, the Graph Convolutional Neural Network (GCNN) is better suited to the problem of predicting the potential energy of an atomistic structure \cite{gu2025high}. \\  

\subsection{Graph Dataset Description}

\begin{figure}[h]
	\centering
	\includegraphics[width=\textwidth]{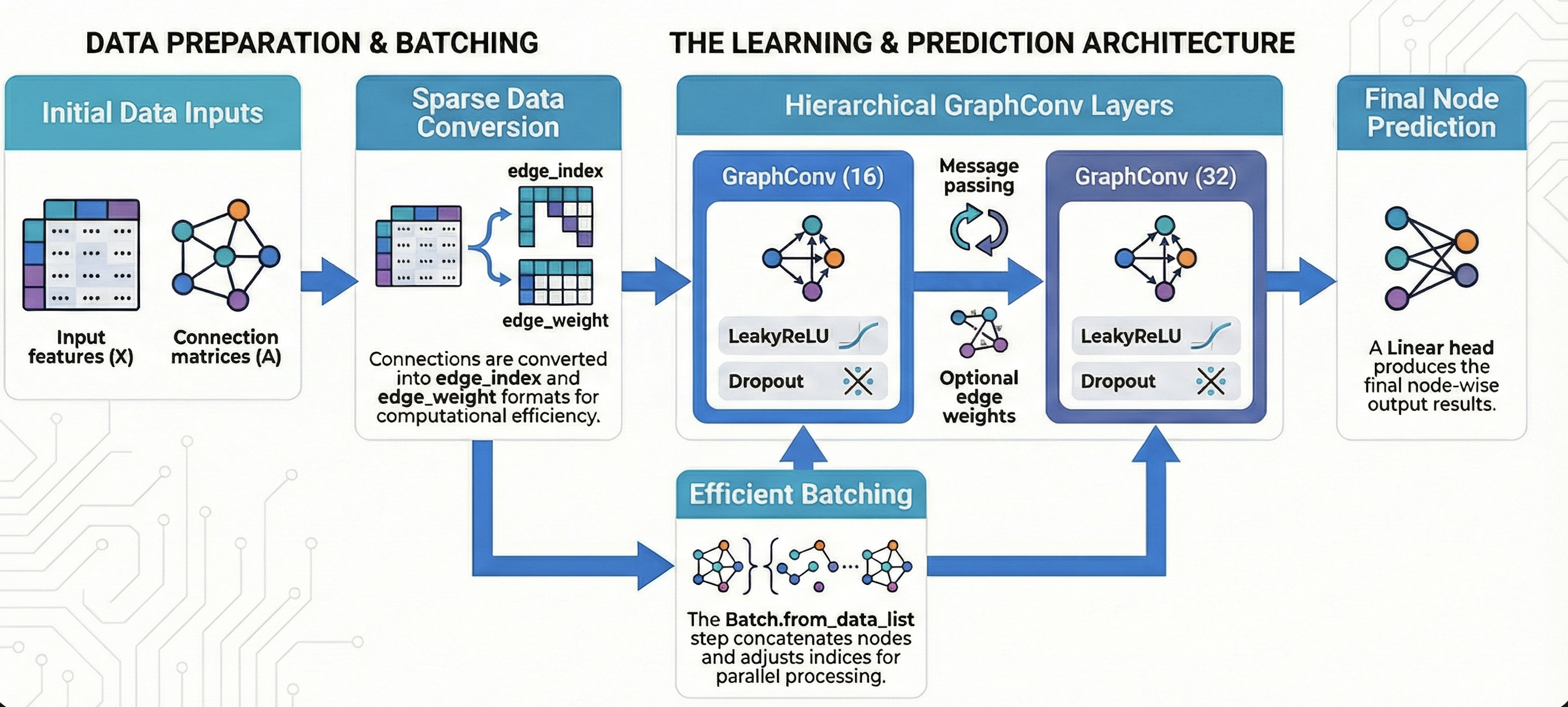}
	\caption{Schematic of the GCNN for \emph{node-level} regression. Inputs $(\mathbf{X},\mathbf{A})$ are converted to sparse tensors; stacked GraphConv layers perform message passing; a linear head produces a scalar at each node. Mini-batches concatenate graphs and adjust indices via \texttt{Batch.from\_data\_list}.}
	\label{fig:gcn}
\end{figure}

\begin{figure}
	\centering
	\begin{subfigure}{0.24\textwidth}
		\centering
		\includegraphics[width=\textwidth]{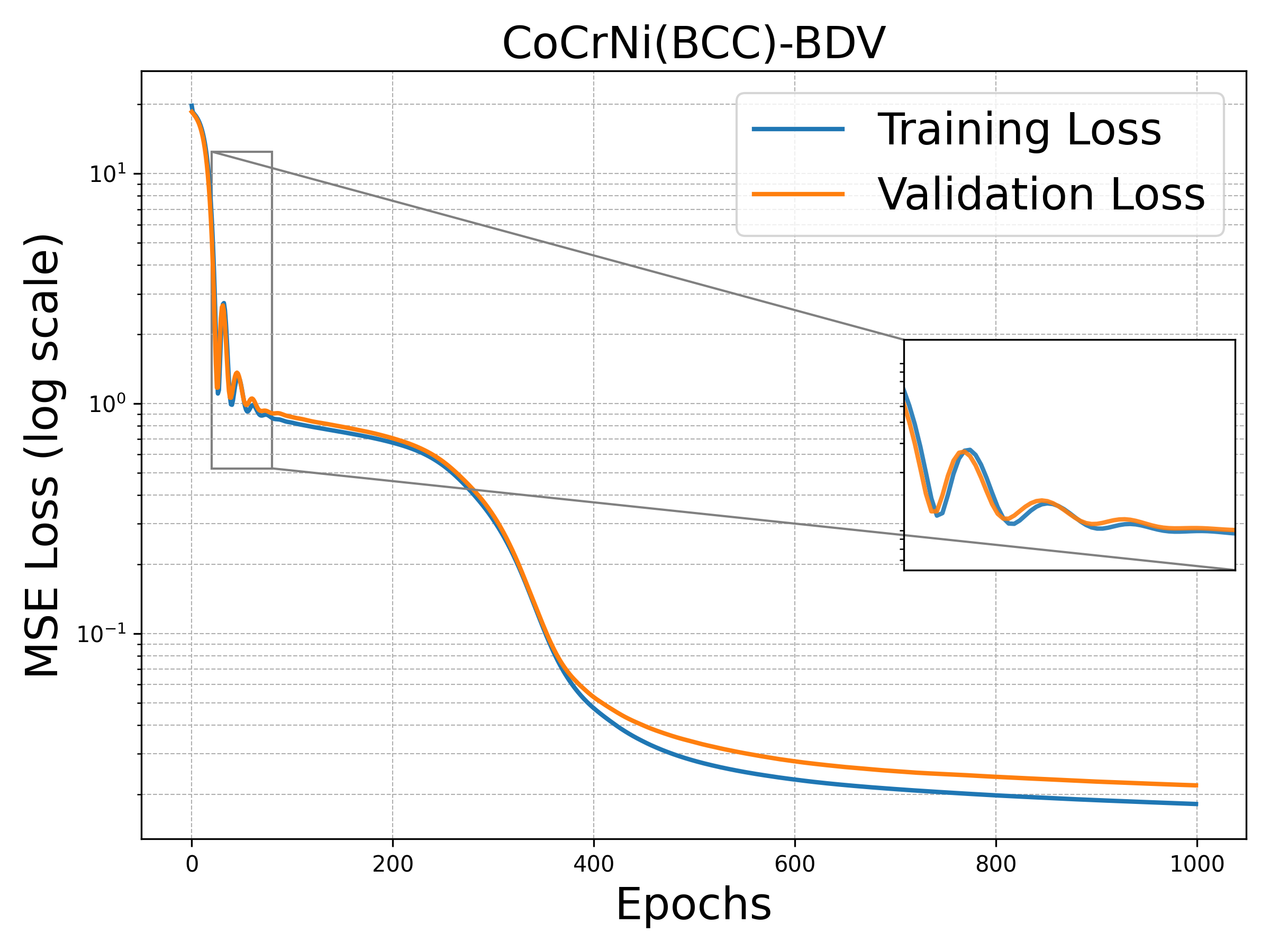}
		\caption{}
		\label{subfig:CCN-bcc-bdv-loss}
	\end{subfigure}
	\begin{subfigure}{0.24\textwidth}
		\centering
		\includegraphics[width=\textwidth]{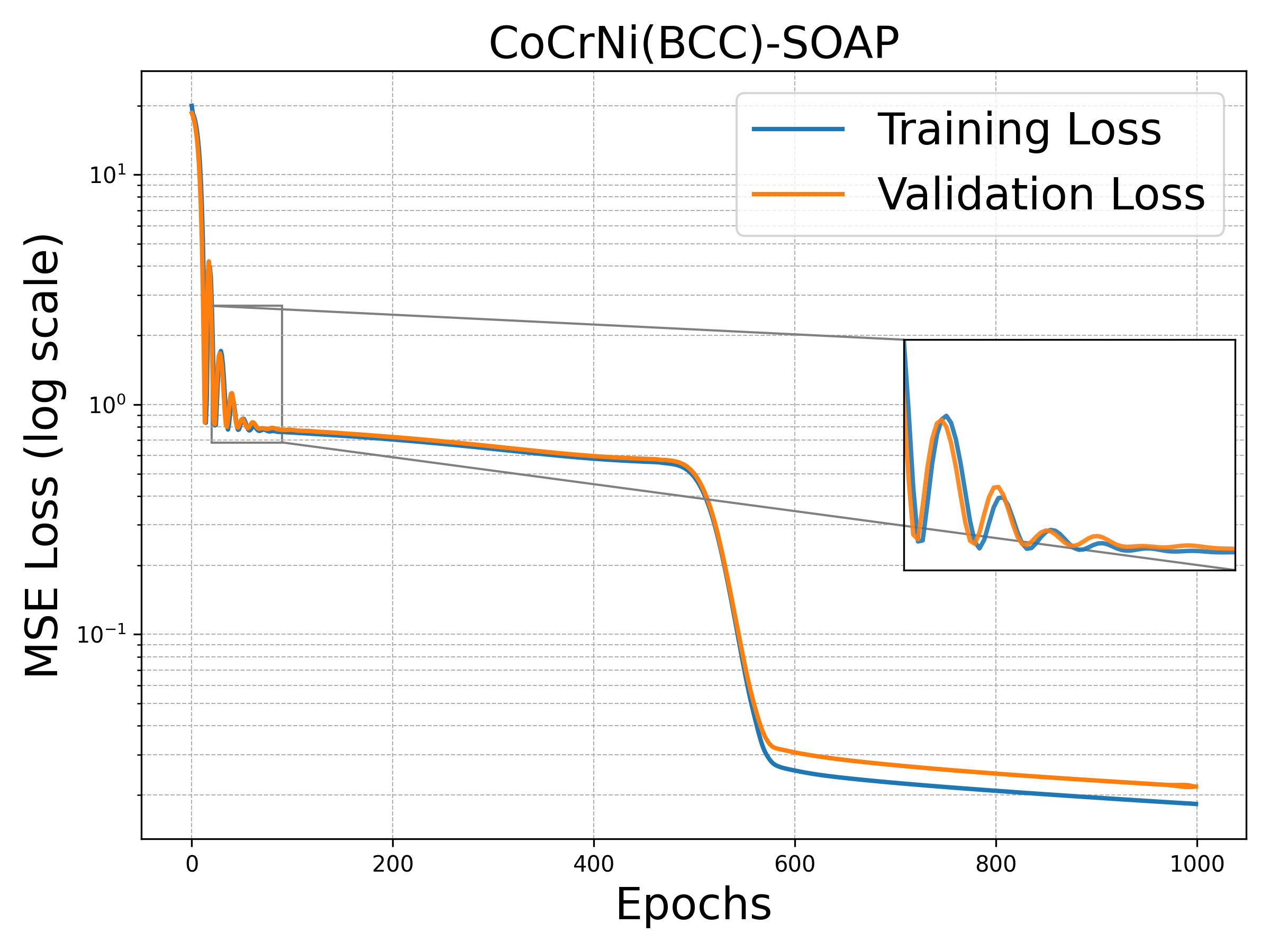}
		\caption{}
		\label{subfig:CCN-bcc-soap-loss}
	\end{subfigure}
	\begin{subfigure}{0.24\textwidth}
		\centering
		\includegraphics[width=\textwidth]{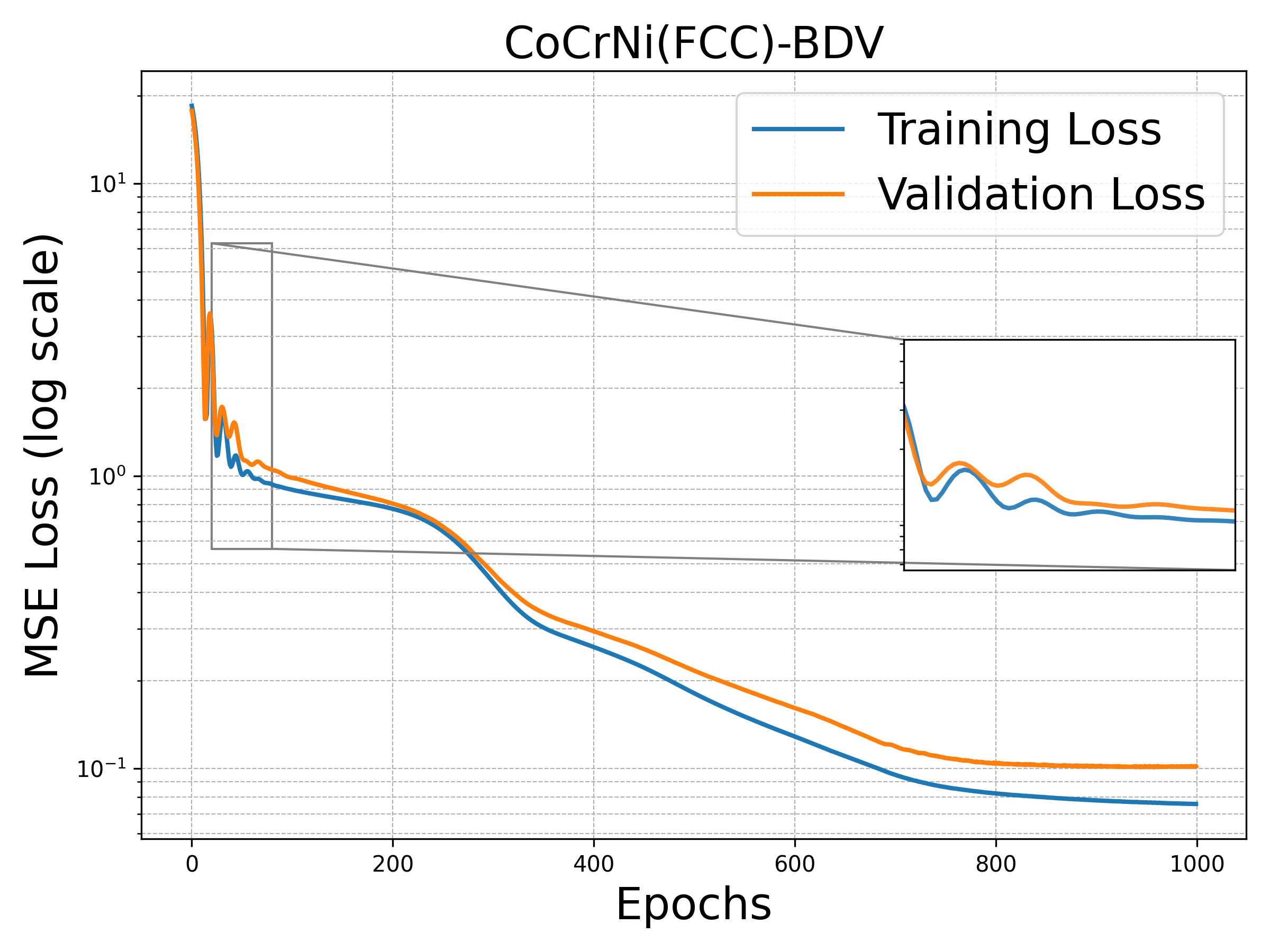}
		\caption{}
		\label{subfig:CCN-fcc-bdv-loss}
	\end{subfigure}
	\begin{subfigure}{0.24\textwidth}
		\centering
		\includegraphics[width=\textwidth]{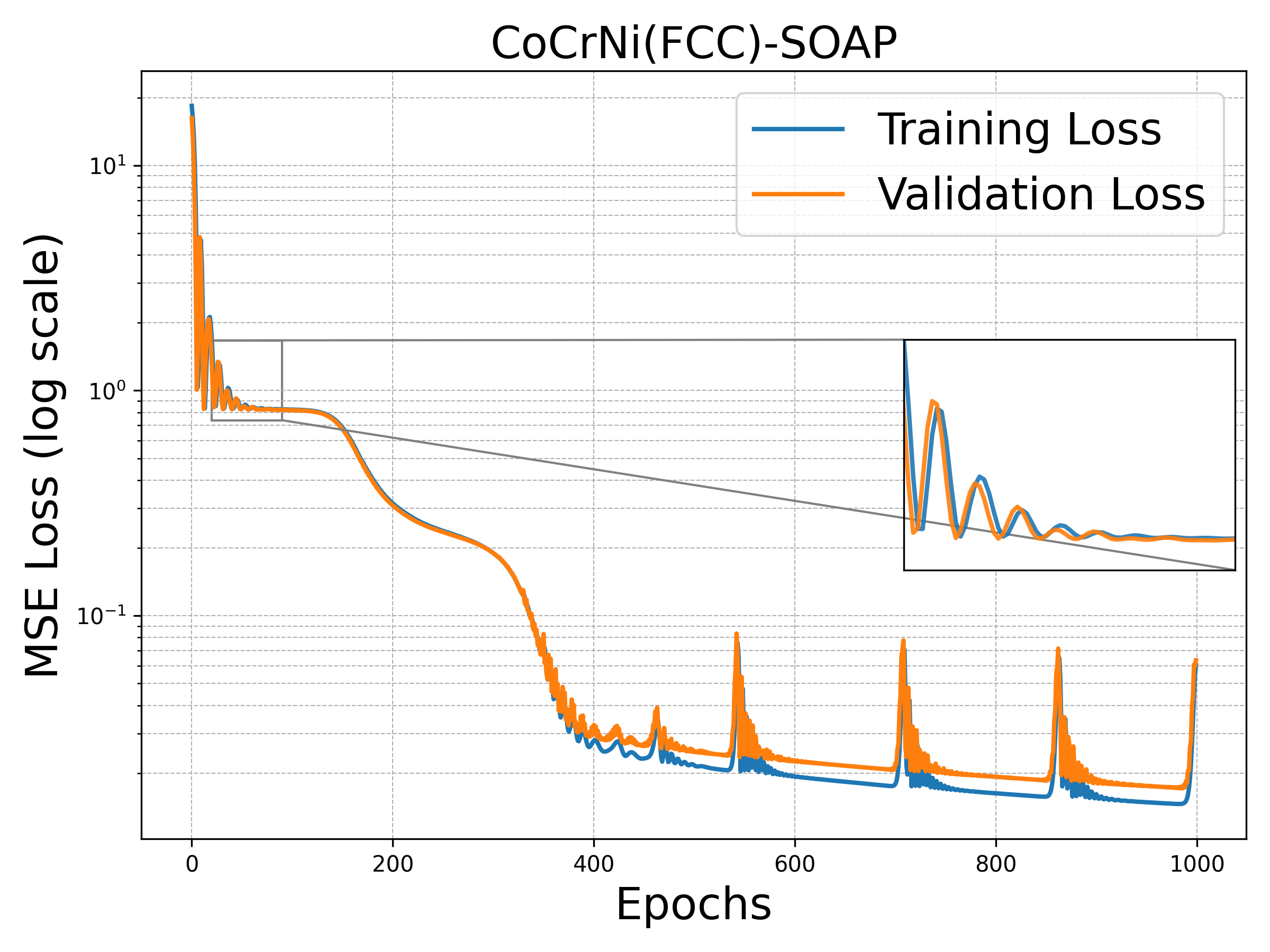}
		\caption{}
		\label{subfig:CCN-fcc-soap-loss}
	\end{subfigure}
\caption{The loss curves for CoCrNi alloy in BCC and FCC crystal structures for BDV (a and c) and SOAP (b and d) descriptors.}
\label{fig:loss}
\end{figure}

A dataset comprising 20 atomic configurations ($\mathrm{(N_g = 20}$)) was used for supervised training of the GCNN. Each configuration corresponds to a distinct atomic structure derived from atomistic simulations. In the present investigation, as proof-of-concept, we have employed the classical interatomic potentials (Embedded Atom Method (EAM) and Modified EAM (MEAM)) for the database generation. However, the proposed framework is extendible to the database generated from \emph{ab initio} calculations and to machine-learned interatomic potentials.
For each configuration  $\emph{i}$, three numerical descriptors were generated. The feature matrix  (\( \mathbf{X}_i \in \mathbb{R}^{N_v \times F} \)) contains the node-level features representing the local atomic environment, derived from the descriptors as defined in section \ref{descriptor} (BDV and SOAP) and featurised by $\mathrm{TF-IDF}$ score (section \ref{feature}). The $\mathrm{N_v}$ and $\mathrm{F}$ is the number of nodes and feature vector size for each node, respectively. The Adjacency matrix (\( \mathbf{A}_i \in \mathbb{R}^{N_v \times N_v} \)) provides the connectivity information constructed from interatomic distances, defining edges between neighbouring atoms within a specified cutoff radius. The adjacency was symmetrised and normalised to ensure an undirected edge representation suitable for message passing. The node-level target vectors  (\( \mathbf{y}_i \in \mathbb{R}^{N_v} \)) are energy-associated node properties derived from the total energies of each configuration. These node-wise energy components served as regression targets for training. Each graph, therefore, represents a single atomic configuration encoded as an undirected, weighted graph  $\mathrm{G_i = (V_i, E_i)}$, where $\mathrm{ |V_i| }$ corresponds to the number of atoms per configuration. The complete dataset was partitioned into training (60\%), validation (20\%), and testing (20\%) subsets. Graphs were batched efficiently using the \texttt{Batch.from\_data\_list()} utility provided by \textsc{PyTorch Geometric}. \\
The predictive model was implemented as a node-level regression GNN using the \texttt{GraphConv} operator. Each convolutional layer updates node embeddings by aggregating information from neighbouring nodes according to:
\[
\mathbf{h}_i^{(l+1)} = \sigma \left( \sum_{j \in \mathcal{N}(i)} w_{ij} \, \mathbf{W}^{(l)} \mathbf{h}_j^{(l)} \right),
\]
where \( \mathbf{h}_i^{(l)} \) is the embedding of node \( i \) at layer \( l \), \( \mathbf{W}^{(l)} \) is a learnable weight matrix, \( w_{ij} \) denotes the edge weight between nodes \( i \) and \( j \), and \( \sigma(\cdot) \) is a LeakyReLU activation function with a negative slope of 0.05. Two graph convolutional layers with hidden dimensions [16, 32] were employed, each followed by dropout regularisation (set to zero in the final production model). The last hidden node embeddings were passed through a linear output layer to produce scalar predictions for each node:
\[
\hat{y}_i = \mathbf{W}_o \mathbf{h}_i^{(L)} + b_o.
\]
No activation was applied to the final layer, enabling the model to capture both positive and negative regression targets (since per-atom energies can be either positive or negative). Model optimisation was performed using the \texttt{AdamW} optimiser with a learning rate of \( 10^{-2} \) and zero weight decay. The mean squared error (MSE) between the predicted and reference node-level values was used as the loss function:
\[
\mathcal{L}_{\mathrm{MSE}} = \frac{1}{N} \sum_{i=1}^{N} \left( \hat{y}_i - y_i \right)^2.
\]
Gradients were backpropagated through all network layers, and gradient clipping with a maximum norm of 1.0 was applied to enhance training stability. The training process was run for multiple epochs until the validation loss converged. The validation and test phases were conducted in evaluation mode, without gradient computation, to assess model generalisation. Both training and evaluation used batched graph data to maximise GPU parallelisation. Table 1 in the SI summarises the parameters and hyperparameters used in GCNN training, and Fig. \ref{fig:gcn} provides the schematic of the GCNN training. All computations were performed in Python~3.10 using \textsc{PyTorch Geometric} (v2.x) on NVIDIA CUDA-enabled hardware. The presented framework is fully modular and easily extensible. The convolution operator can be substituted with \texttt{GCNConv}, \texttt{SAGEConv}, or \texttt{GATConv} without altering data structures. The architecture can be extended to graph-level prediction tasks using pooling operations such as \texttt{global\_mean\_pool} or \texttt{global\_add\_pool}. Moreover, inclusion of physically meaningful edge weights, such as bond distances, angles, or coordination strengths, enables integration of domain knowledge into the learning process.
The Fig. \ref{fig:loss} shows the variation of the training and validation loss for the BCC and FCC structure of CoCrNi alloy. The evolution of the training and validation losses during optimization reveals clear differences between the two descriptors. For the SOAP descriptor, the loss initially decreases rapidly and then exhibits pronounced oscillations and intermittent sharp reductions before gradually converging to a lower error level. These oscillatory features indicate periods where the optimizer temporarily departs from a local minimum and subsequently re-enters a region of improved convergence. In contrast, the BDV descriptor shows a smoother and more monotonic decline in both training and validation loss. After the initial rapid decrease, the loss decreases steadily with minimal fluctuations, eventually approaching a stable plateau at a slightly higher error than that obtained with SOAP. In both cases, the training and validation curves remain close throughout the optimization process, suggesting that overfitting is limited and that the models generalize well within the available dataset. The contrasting convergence patterns, therefore, primarily reflect differences in the underlying descriptor representations rather than deficiencies in the learning procedure. The distinct training behaviours observed for the SOAP and BDV descriptors can be understood in terms of the geometry of the loss landscape induced by the descriptor representations. The model parameters $\theta$ are obtained by minimizing the mean squared error loss

\begin{equation}
\mathrm{
L(\theta) = \frac{1}{N}\sum_{i=1}^{N}\left(y_i - f_{\theta}(x_i)\right)^2 ,
}
\end{equation}
where $\mathrm{x_i}$ denotes the descriptor of the local atomic environment and $\mathrm{f_{\theta}}$ is the graph neural network mapping. The optimization dynamics are governed by the curvature of the loss surface, characterized by the Hessian matrix
\begin{equation}
\mathrm{
H_{ij} = \frac{\partial^2 L}{\partial \theta_i \, \partial \theta_j}.
}
\end{equation}
For the simpler BDV descriptor, which has a relatively low-dimensional feature space and weak inter-feature correlations, the eigenvalue spectrum of the Hessian is narrow, resulting in a smoother loss surface and a more stable gradient descent trajectory with nearly monotonic convergence. In contrast, the SOAP descriptor comprises a high-dimensional expansion of radial and angular basis functions, leading to stronger feature correlations and a broader Hessian eigenvalue spectrum with large curvature variations. Consequently, the optimisation landscape contains narrow valleys and sharper minima, so that for a finite learning rate the gradient updates may temporarily overshoot the minimum along directions associated with large eigenvalues, producing the oscillatory behaviour and sudden reductions in loss observed during training. This mathematical distinction reflects the trade-off between the expressiveness of descriptors and the complexity of optimisation in graph-based learning of atomic environments. The SI summarises the loss curves for the other alloys (see Fig. 3-11 in the SI).

\begin{figure}
	\centering
	\begin{subfigure}{\textwidth}
		\centering
		\includegraphics[width=\textwidth]{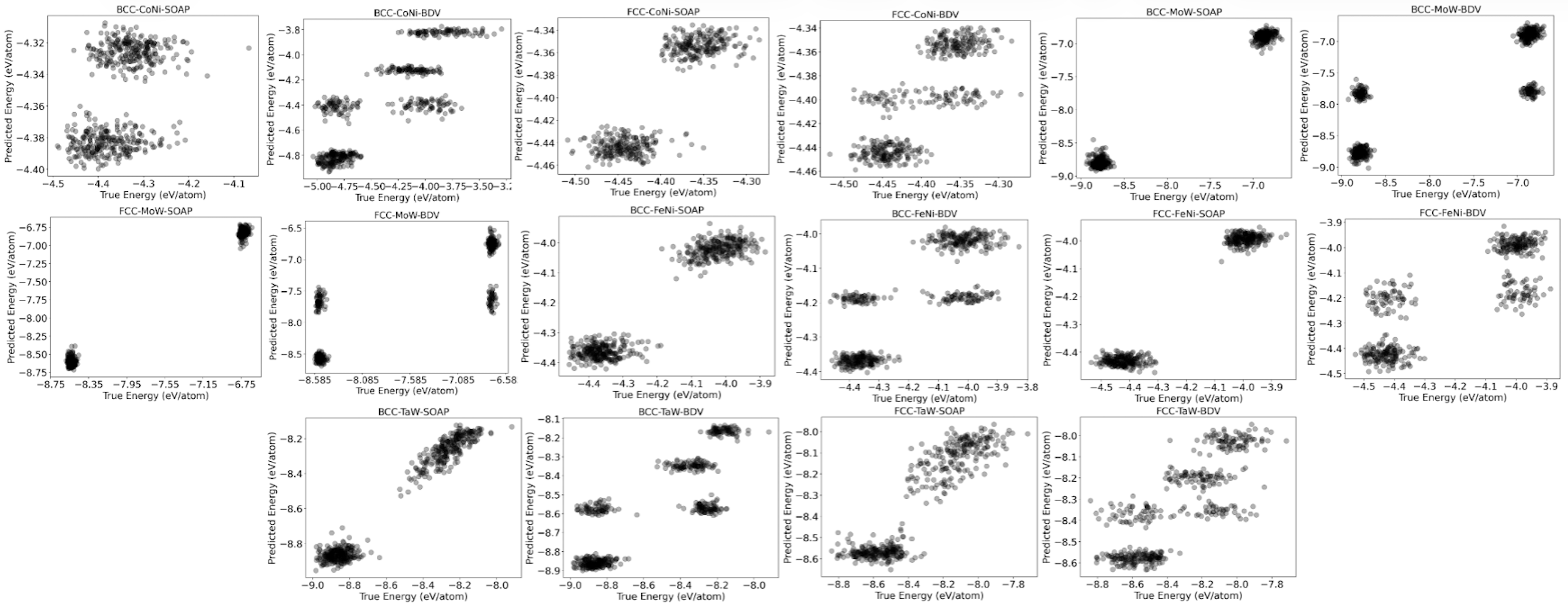}
		\caption{}
		\label{subfig:binary-energy}
	\end{subfigure}
	\begin{subfigure}{0.65\textwidth}
		\centering
		\includegraphics[width=\textwidth]{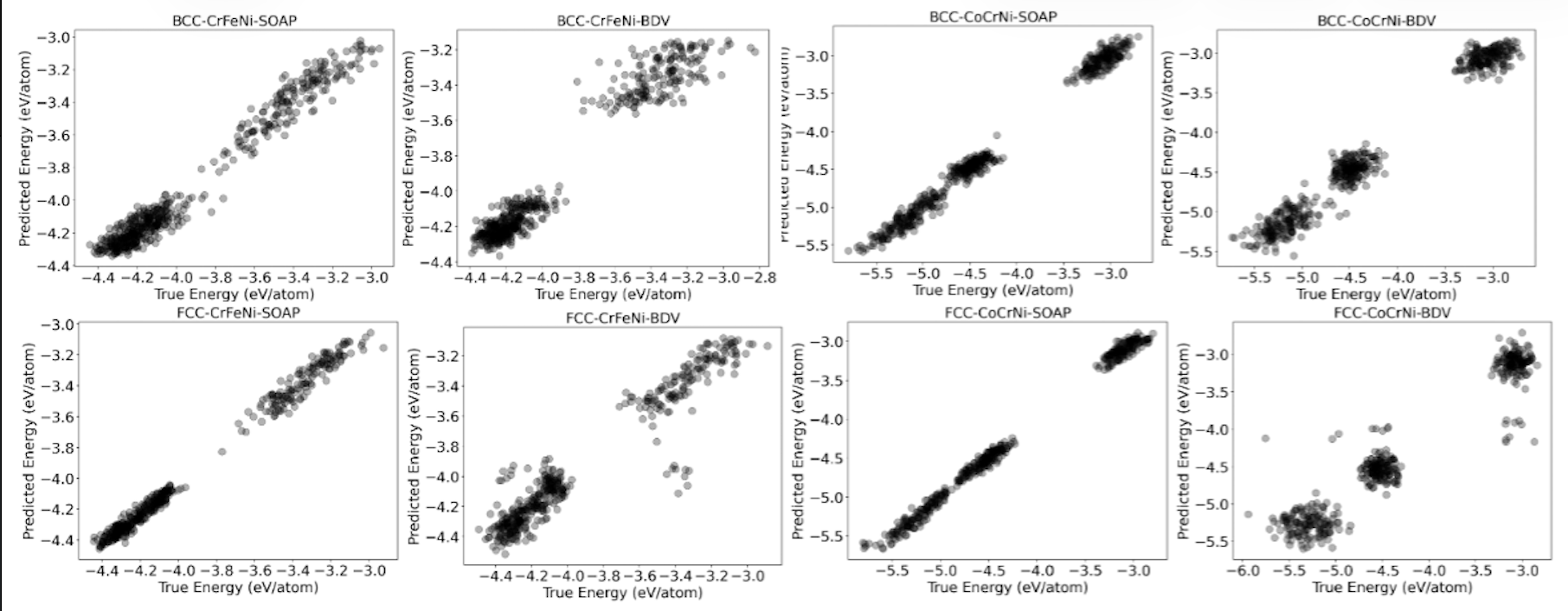}
		\caption{}
		\label{subfig:ternary-energy}
	\end{subfigure}
	\begin{subfigure}{0.34\textwidth}
		\centering
		\includegraphics[width=\textwidth]{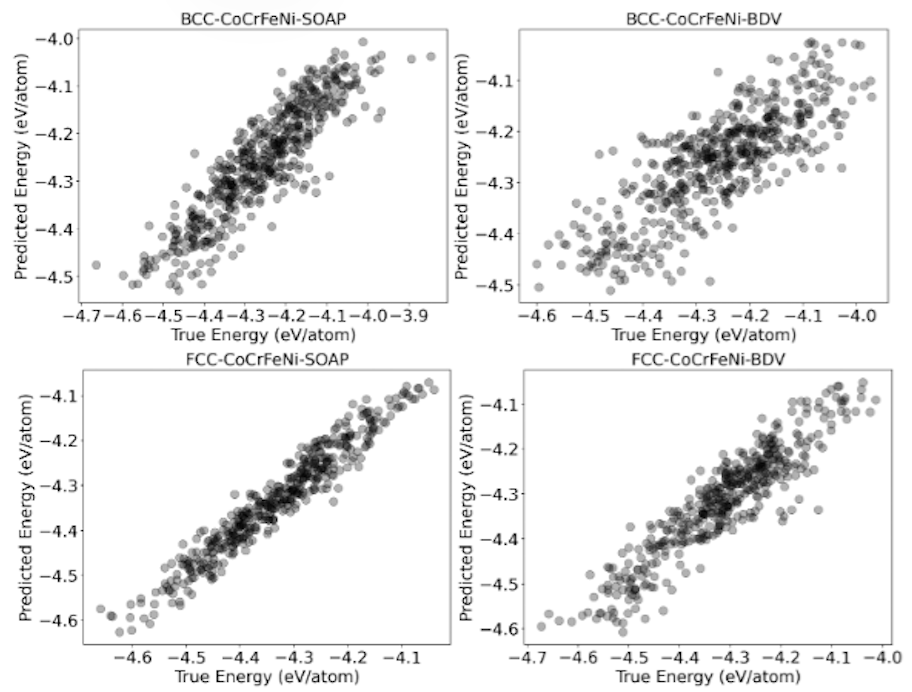}
		\caption{}
		\label{subfig:quaternary-energy}
	\end{subfigure}
	\begin{subfigure}{\textwidth}
		\centering
		\includegraphics[width=\textwidth]{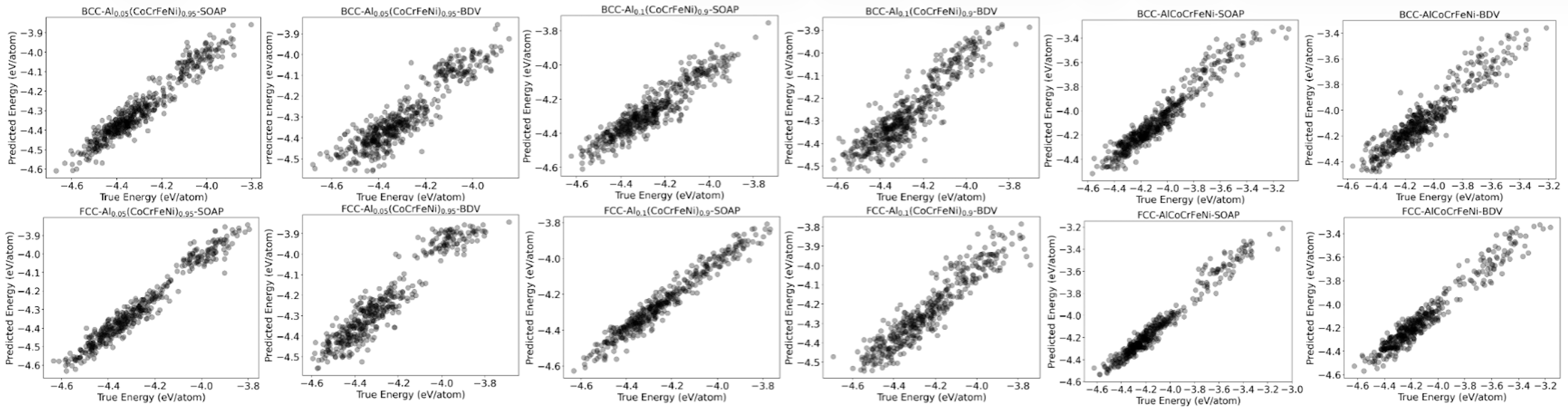}
		\caption{}
		\label{subfig:quinary-energy}
	\end{subfigure}
\caption{The energy prediction curves for (a) binary, (b) ternary, (c) quaternary and (d) quinary alloys}
\label{fig:energy}
\end{figure}

\begin{figure}[h]
	\centering
	\begin{subfigure}{0.49\textwidth}
		\centering
		\includegraphics[width=\textwidth]{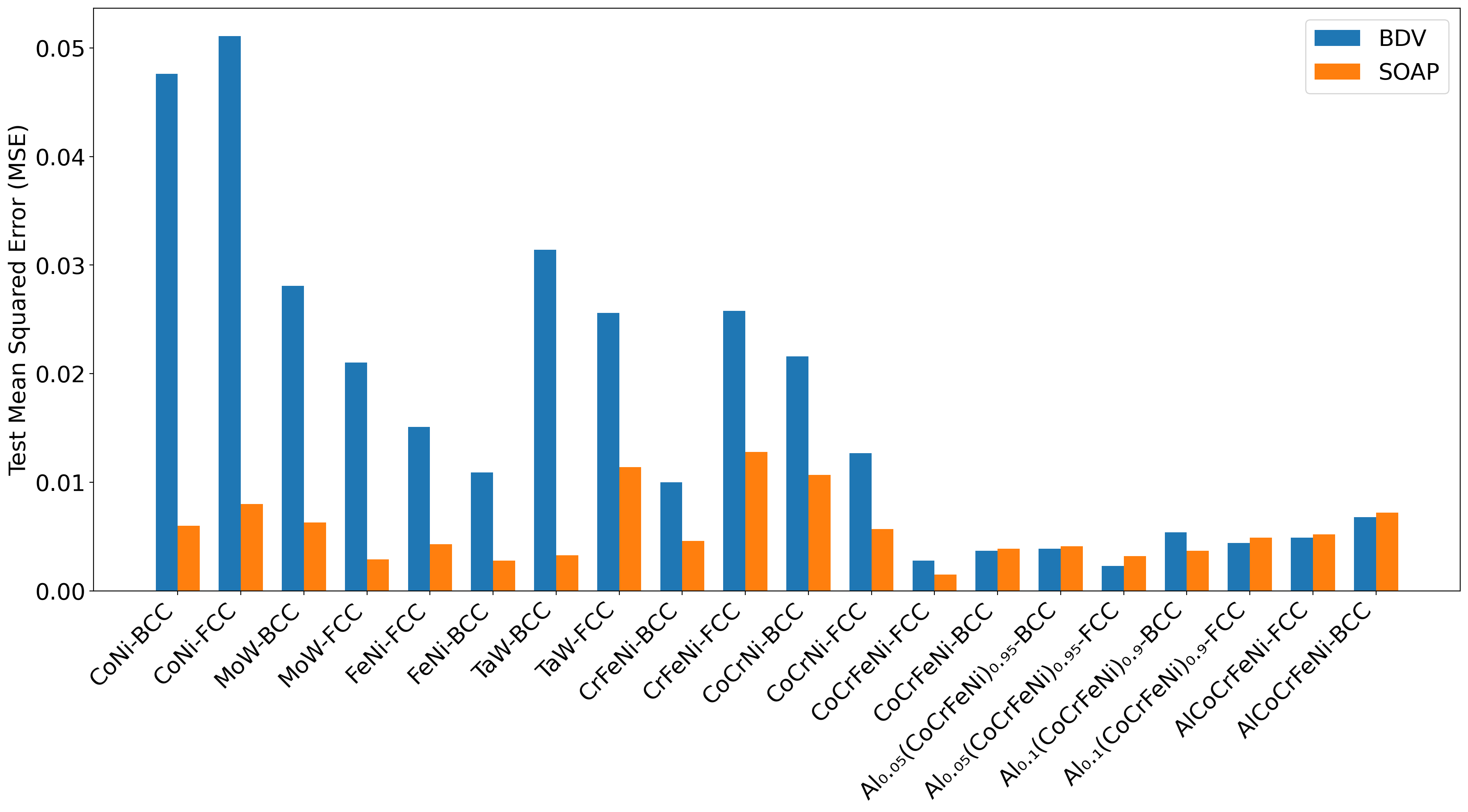}
		\caption{a}
		\label{subfig:mse}
	\end{subfigure}
	\begin{subfigure}{0.49\textwidth}
		\centering
		\includegraphics[width=\textwidth]{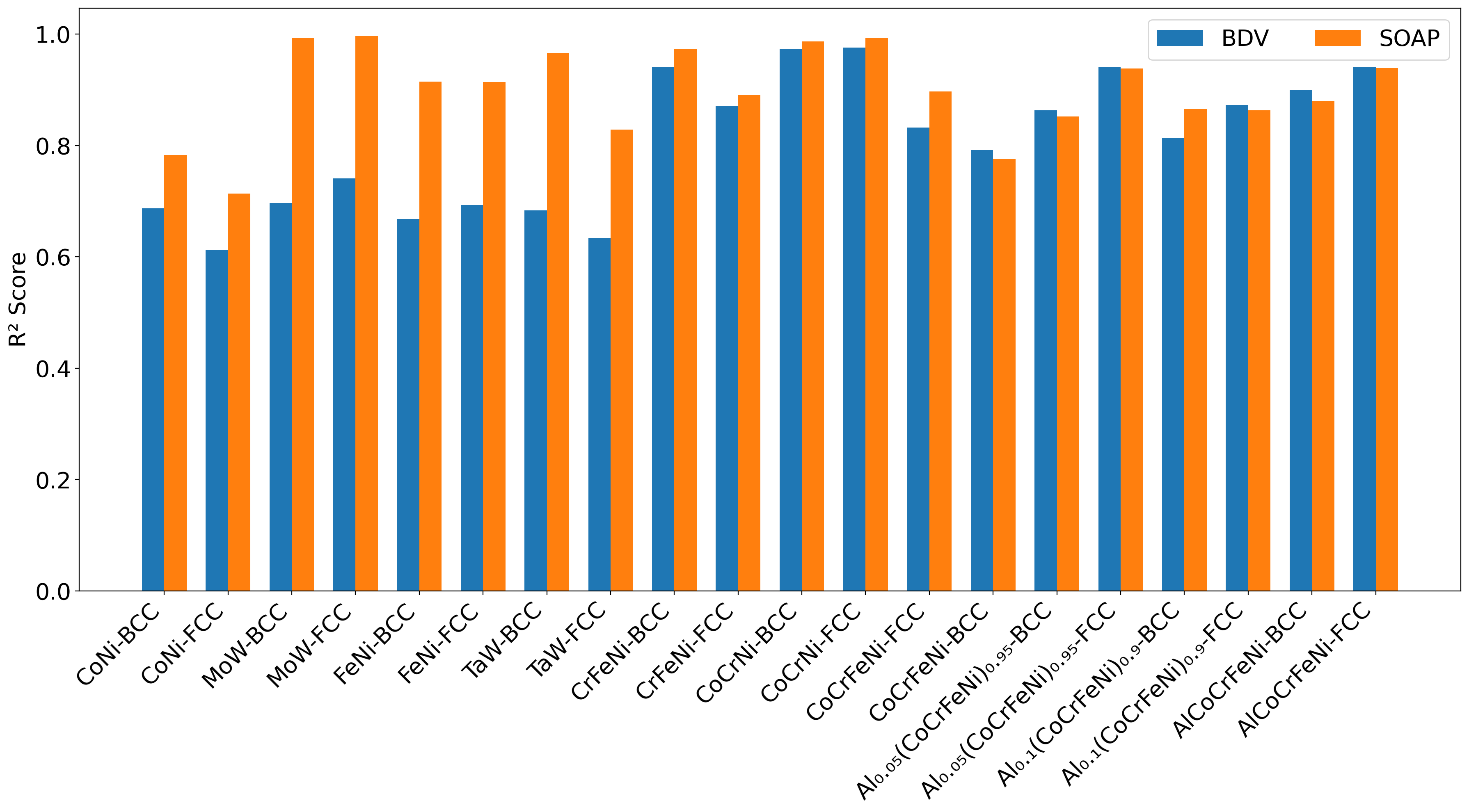}
		\caption{a}
		\label{subfig:rsquare}
	\end{subfigure}
	\caption{Comparison of the performance with (a) test MSE and (b) $\mathrm{R^2}$ score of BDV and SOAP as descriptors for binary, ternary, quaternary, and quinary alloys}
	\label{fig:r2mse}
	
\end{figure}
\subsection{Alchemical Monte Carlo based Sampling}
The combinatorial sampling of the alloys in the candidate crystal structure is carried out using the Genetic Algorithm-based Atomistic Sampling Protocol (GAASP) \cite{anand2023gaasp}. The GAASP approach employs the alchemical swap with Metropolis acceptance criteria to perform biased sampling (Algorithm \ref{alg:GAASP}). In the present investigation, we explored the low-energy thermodynamically relevant atomistic configurations of alloys in the candidate crystal structures.  The alchemical Monte Carlo-based sampling ensures efficient atomistic sampling and holistic exploration of the potential energy landscape. In the present work, the alchemical Monte Carlo swaps atomic positions, while the GCNN-based model predicts the configuration's potential energy. The GCNN-based ML model is trained with both SOAP and BDV descriptors. 

\begin{figure*}[h]
	\centering
	\begin{subfigure}{0.45\textwidth}
		\centering
		\includegraphics[width=\textwidth]{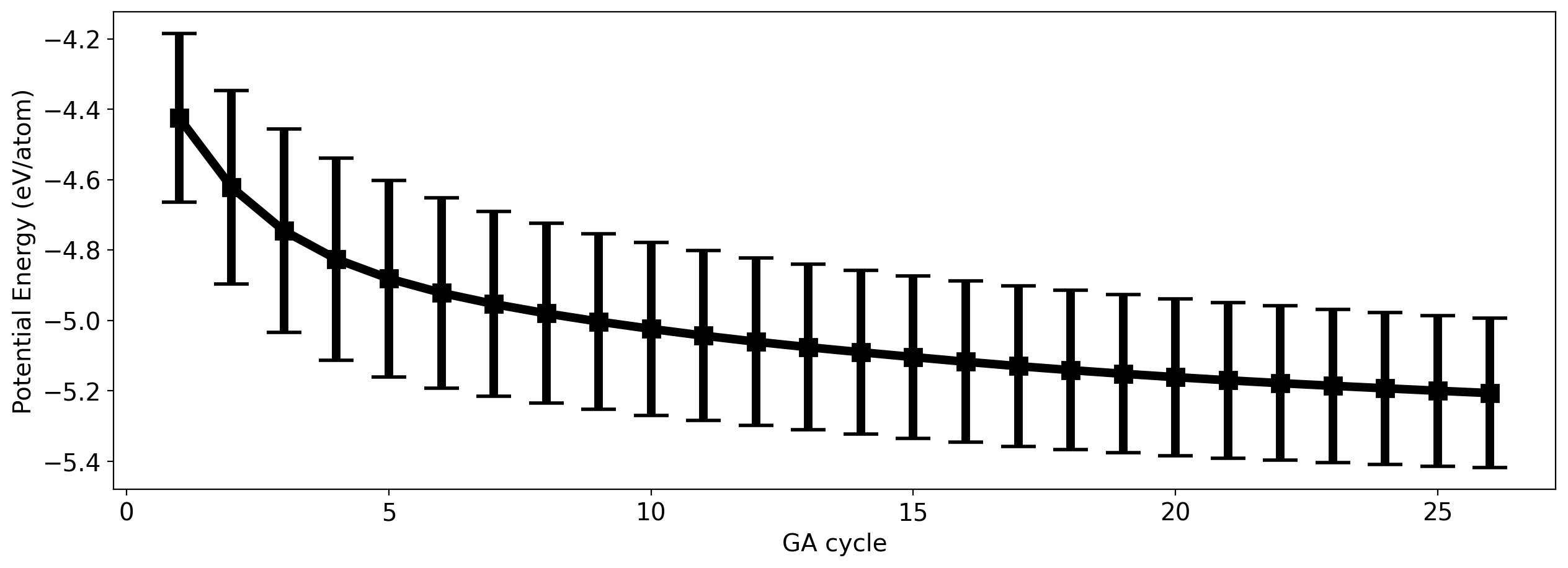}
		\includegraphics[width=\textwidth]{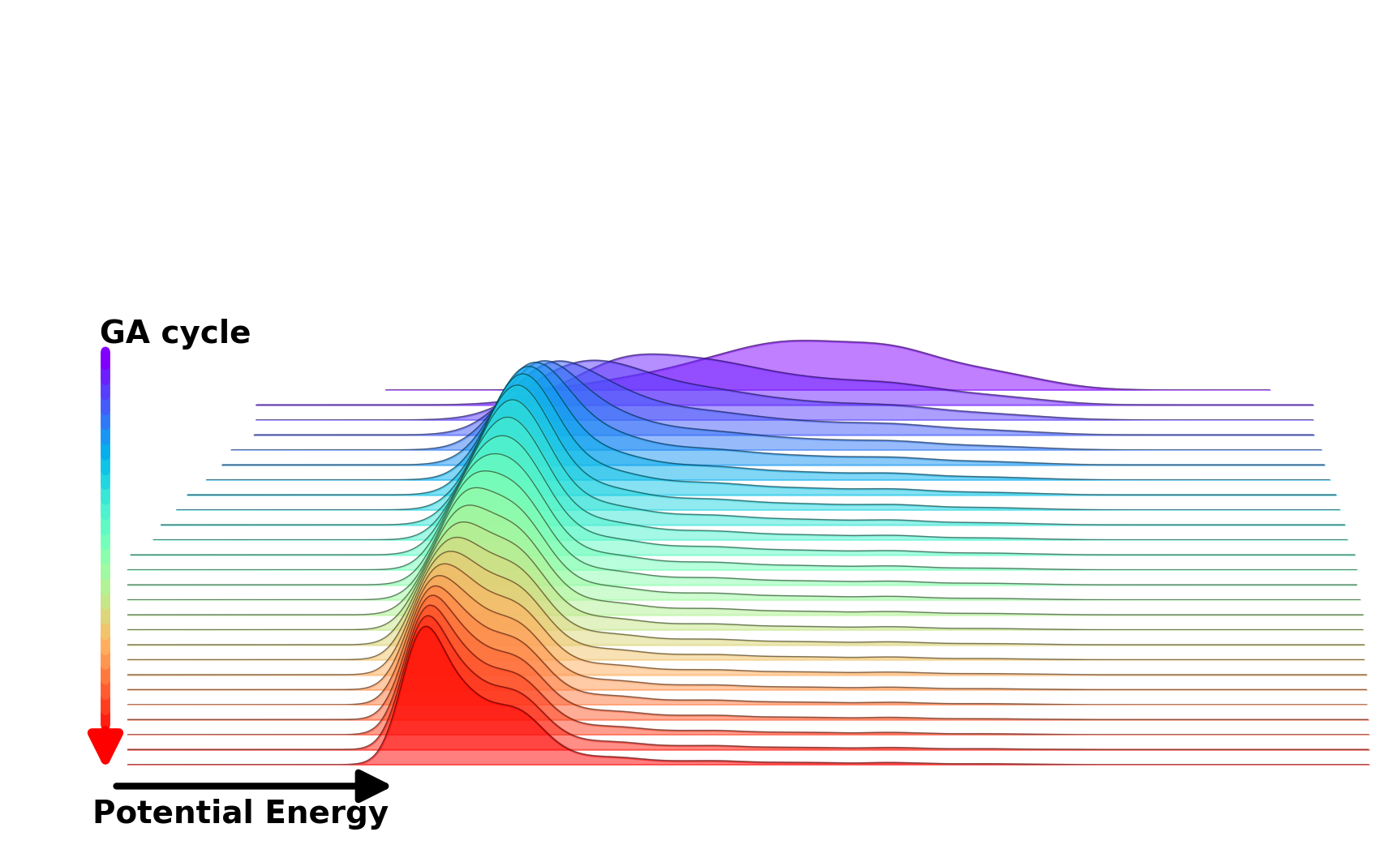}
		\caption{}
		\label{subfig:meanenergyBCC}
	\end{subfigure}
	\begin{subfigure}{0.45\textwidth}
		\centering
		\includegraphics[width=\textwidth]{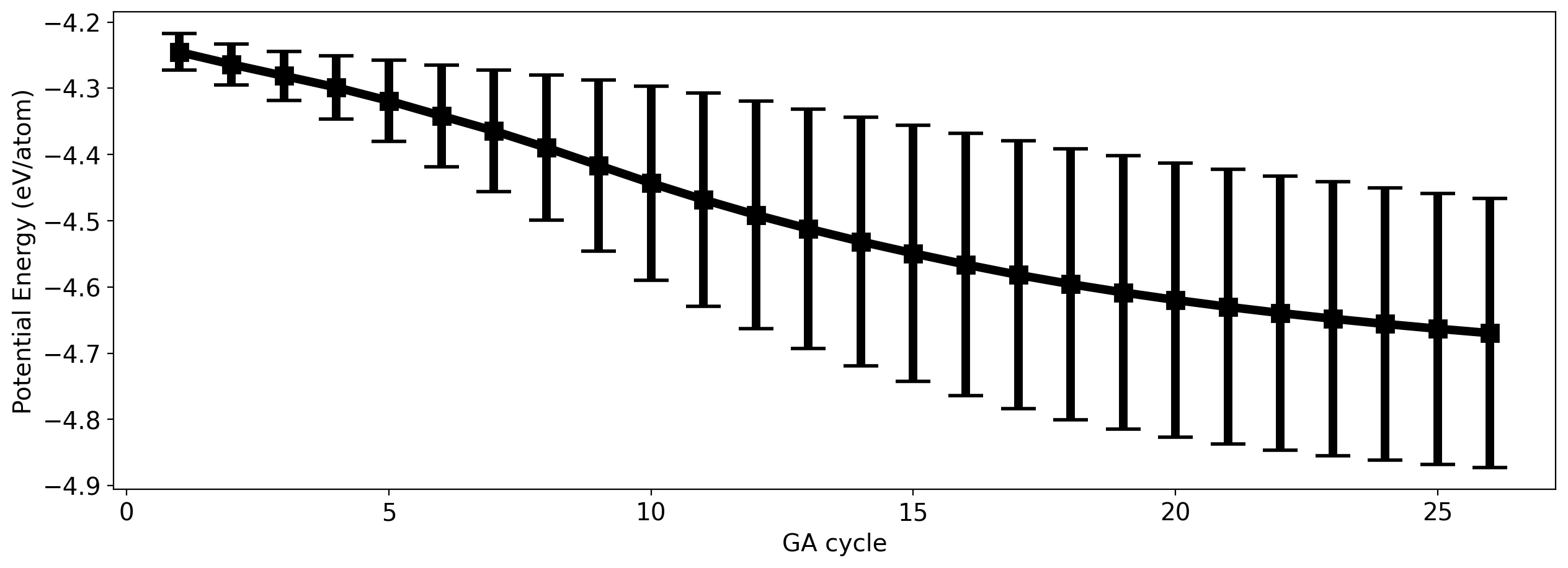}
		\includegraphics[width=\textwidth]{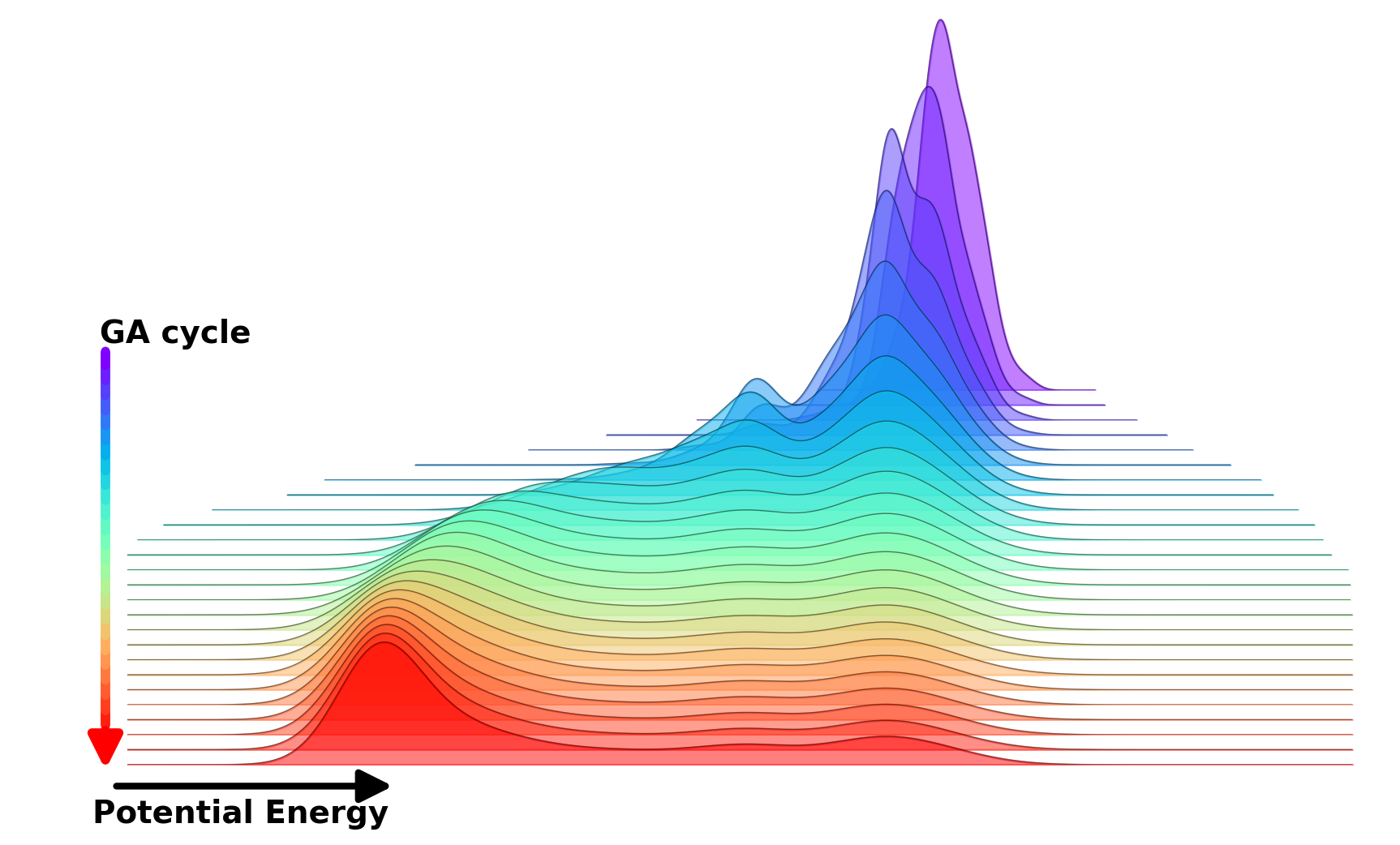}
		\caption{}
		\label{subfig:meanenergyFCC}
	\end{subfigure}
	\caption{The variation in the mean energy and corresponding energy distribution evolution with alchemical Monte Carlo swap with GCNN ML model for (a) BCC and (b) FCC crystal structure of $\mathrm{Al_{0.05}(CoCrFeNi)_{0.95}}$ high-entropy alloy}
	\label{fig:energytrend}
\end{figure*}

\begin{tcolorbox}[colback=white,colframe=black,title=Algorithm 1- GAASP: Genetic Algorithm-Based Atomistic Sampling Protocol]

\textbf{Input:} Supercell with $\mathrm{N_A}$ atoms; population size $N$; GA cycles $G$; number of parents $X$; temperature $T$

\textbf{Output:} Sampled atomistic configurations representing the energy landscape

\begin{algorithm}[H]
\DontPrintSemicolon
\label{alg:GAASP}

Generate $\mathrm{N}$ random atomistic configurations\;

Compute configurational energy of each configuration\;

\For{$g=1$ \KwTo $G$}{

Sort configurations by increasing energy\;

Select $\mathrm{X}$ parents using roulette-wheel selection\;

Pair parents and encode atomic configurations\;

Perform alchemical swaps between parents\;

Repeat swaps $\mathrm{Q}$ times where $Q \in [0.1N_A,0.3N_A]$\;

Compute energy ($\mathrm{E}$) of generated children with GCNN ML model\;

Accept children with probability ($\mathrm{P}$) using the Metropolis criterion  (if $\mathrm{\Delta E > 0}$)
$\mathrm{P = \exp(-\Delta E/k_B T)}$\;

Update population with accepted configurations\;

}

Return sampled configurations\;

\end{algorithm}

\end{tcolorbox}
The Fig. \ref{fig:energytrend} shows the energy trend with the alchemical Monte Carlo cycle with the GAASP approach for the BCC and FCC crystal structures of $\mathrm{Al_{0.05}(CoCrFeNi)_{0.95}}$ high-entropy alloy. In the early GA cycles, the population exhibits a broad distribution of relatively high-energy configurations, indicating that the algorithm samples a wide variety of structural arrangements within the configurational space of the specific crystal structure (BCC or FCC) considered in each independent GA exploration. As the GA progresses, selection and crossover gradually favour structures with lower potential energy, shifting the distribution toward the lower-energy region of the landscape. For the BCC environment, the population rapidly concentrates around a stable basin, reflecting the identification of energetically favourable BCC-like configurations and a relatively smooth descent toward the minimum. In contrast, the FCC case exhibits a broader and more structured distribution during intermediate cycles, suggesting that the GA explores several competing local minima before converging. This behaviour indicates that the descriptor representation captures subtle variations in atomic environments, allowing the GA to navigate multiple metastable configurations and progressively refine the population toward increasingly stable crystal arrangements. The accompanying decrease in the mean potential energy with GA cycles further confirms the systematic movement of the population toward deeper regions of the potential energy surface corresponding to stable BCC and FCC structural motifs.
\section{Results and discussions}
\subsection{Comparison of BDV and SOAP descriptors for predicting the potential energy of binary, ternary, quaternary and quinary alloys}
Figure \ref{subfig:binary-energy}, \ref{subfig:ternary-energy}, \ref{subfig:quaternary-energy}, and \ref{subfig:quinary-energy} show the performance of the GCNN ML model with both the BDV and SOAP descriptors for binary (CoNi, MoW, FeNi, and TaW), ternary (CoCrNi and CrFeNi), quaternary (CoCrFeNi) and quinary  ($\mathrm{Al_x(CoCrFeNi)_{1-x}}$ with $\mathrm{x = 0.05, 0.1 and 0.2}$), respectively in BCC and FCC crystal structures. The performance of the GCNN for the binary alloys is particularly interesting, as the real per-atom energies are segregated into two bands for each atomic species, and the GCNN model with BDV descriptors tends to predict the mean energies for both atom types, which is a classical issue with low-information descriptors, while the SOAP-based GCNN ML model does not exhibit such a limitation. As the compositional complexity of the alloy increases, \emph{i.e.,} from ternary to quinary chemically disordered alloys, the energy landscape becomes more diffuse, and the performance of the BDV descriptor becomes interestingly comparable to that of SOAP. Fig. \ref{subfig:mse} and \ref{subfig:rsquare} show the MSE and $\mathrm{R^2}$ for all alloys, comparing the performance of the BDV and SOAP descriptors. A trend of the simple descriptor (\emph{i.e.,} BDV) is exhibiting comparable performance to the complex descriptor (\emph{i.e.,} SOAP) for the complex alloys, while the complex descriptor for simpler alloys (\emph{i.e.,} binary) shows better performance. Such an observation is crucial for the descriptor design, as the potential energy landscape of the materials needs to be explored to inform better decisions.     

\subsection{Information entropy and its application for the prediction of phases among the candidate crystal structures}
\begin{figure}[h]
	\centering
	\begin{subfigure}{0.3\textwidth}
		\centering
		\includegraphics[width=\textwidth]{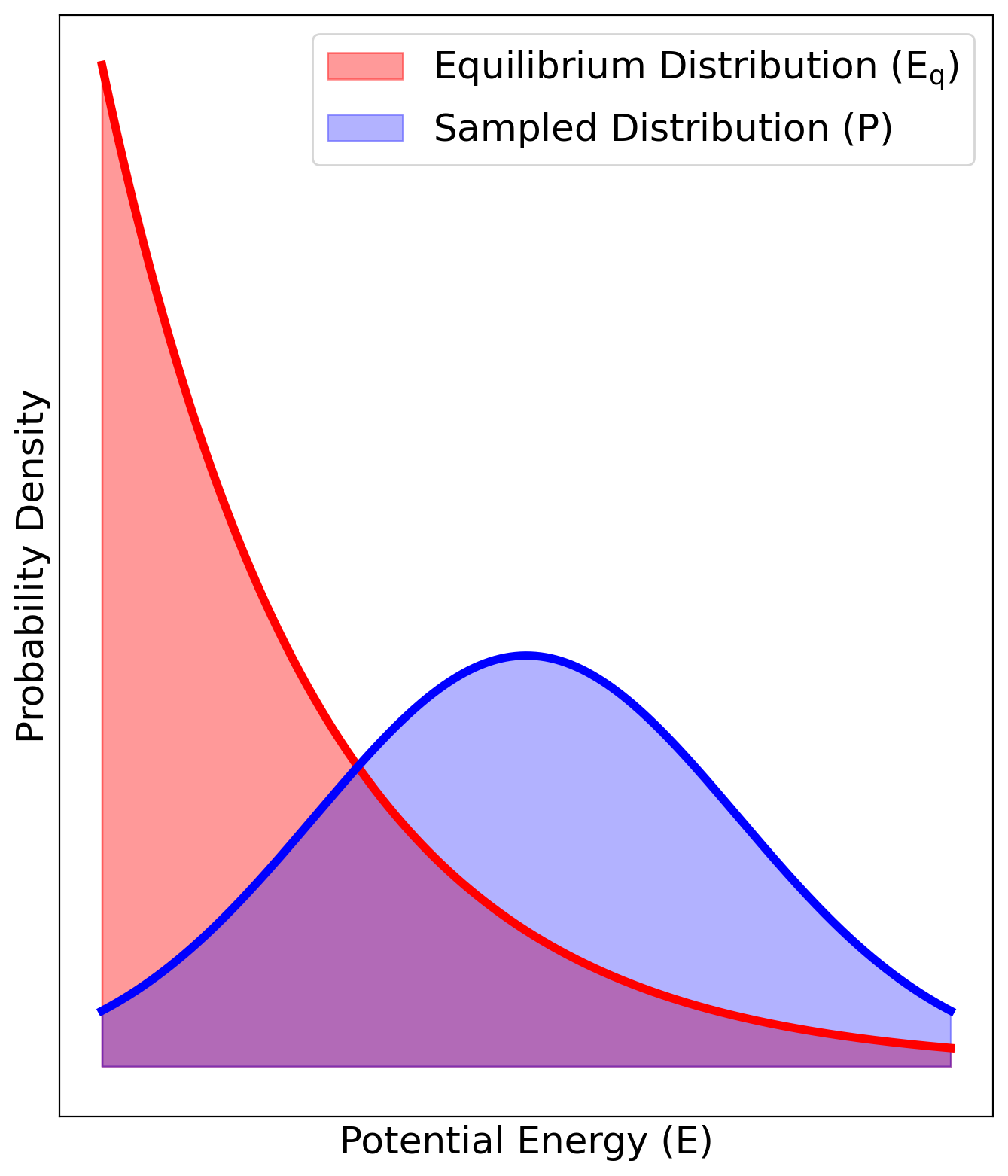}
		\caption{a}
		\label{subfig:pdfall}
	\end{subfigure}
	\begin{subfigure}{0.31\textwidth}
		\centering
		\includegraphics[width=\textwidth]{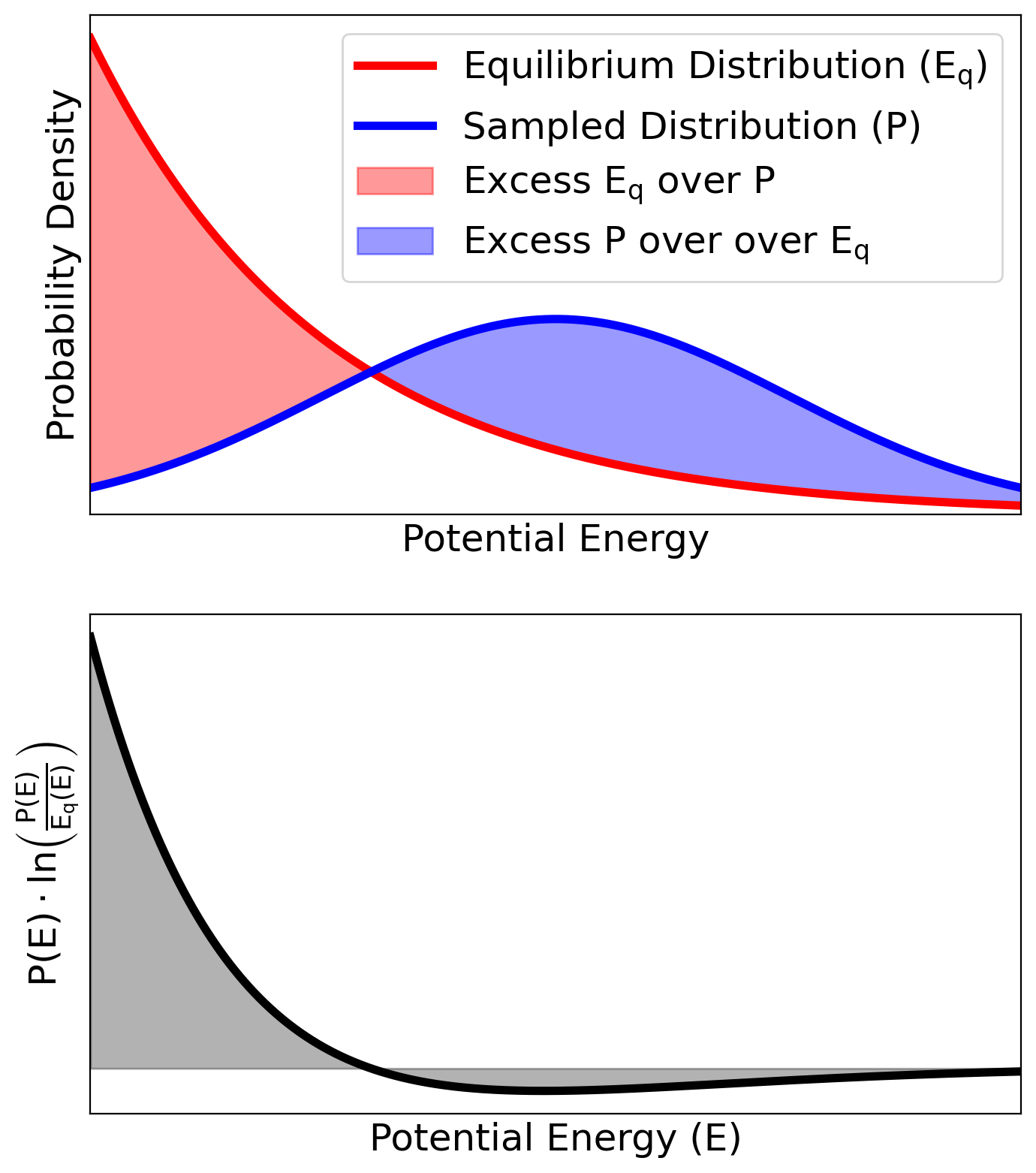}
		\caption{}
		\label{subfig:kld}
	\end{subfigure}
	\begin{subfigure}{0.29\textwidth}
		\centering
		\includegraphics[width=\textwidth]{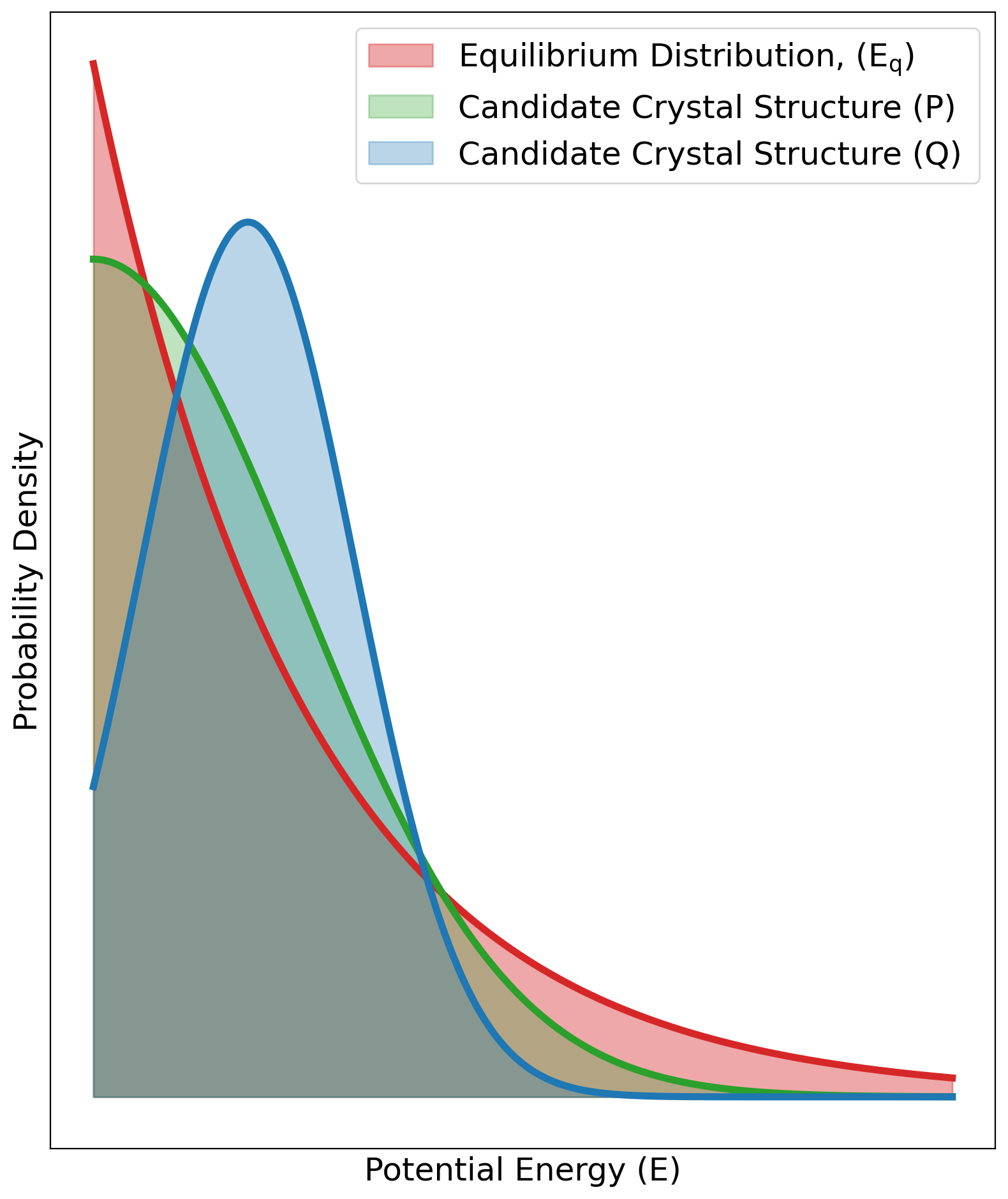}
		\caption{}
		\label{subfig:unbound}
	\end{subfigure}
	\caption{(a) The description of Kullback-Leibler (KL) divergence $\mathrm{D_{KL}}$ between the sampled potential energy landscape (P) and reference or equilibrium potential energy landscape, (b) The graphical explanation of the $\mathrm{D_{KL}}$ value, and (c) schematic representation of the crystal structure prediction problem as an information alignment problem.}
	\label{fig:schmatic-kld}
	
\end{figure}

\begin{figure*}[!ht]
	\centering
	\begin{subfigure}{\textwidth}
		\centering
		\includegraphics[width=0.8\textwidth]{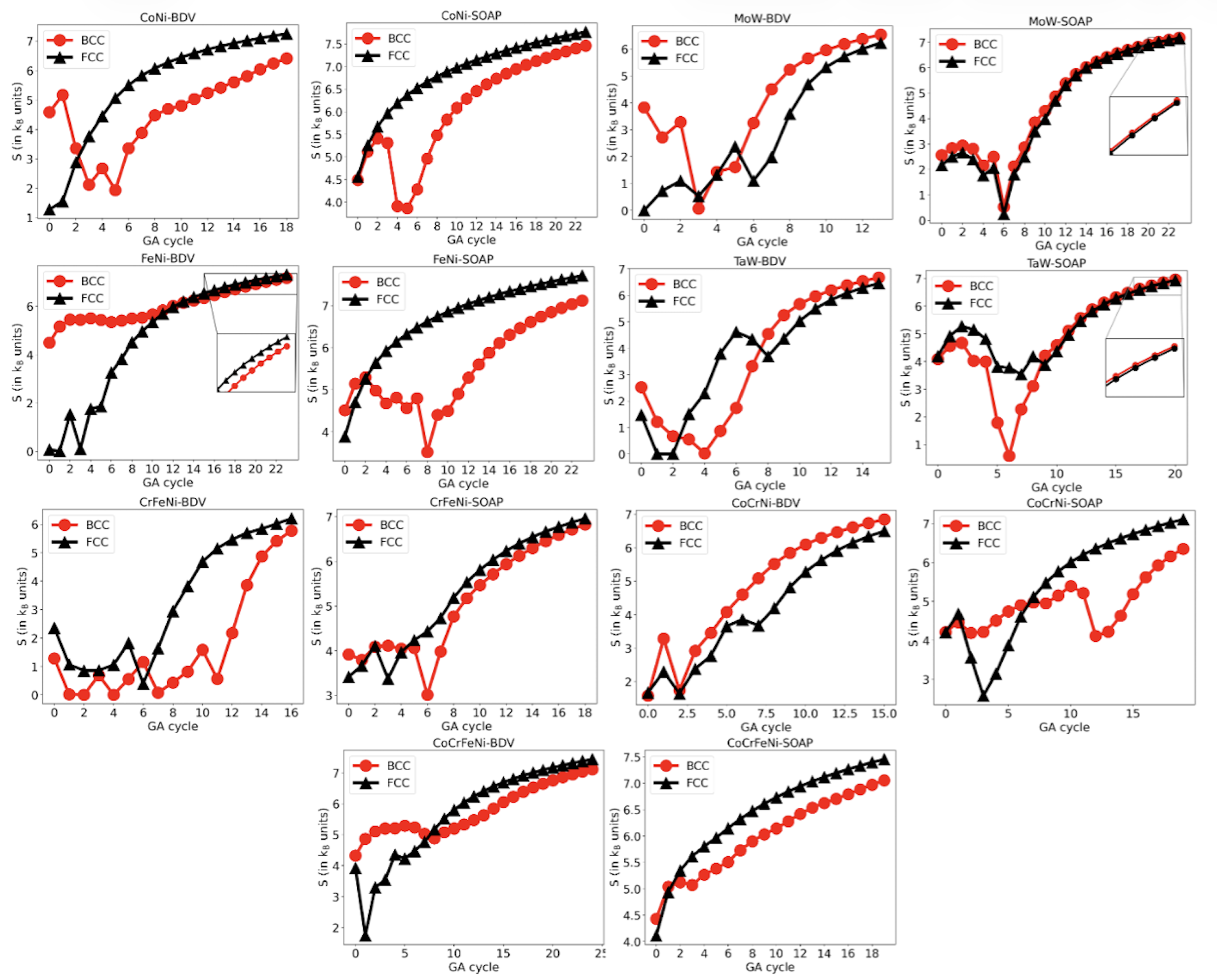}
		\caption{}
		\label{subfig:soapbdv-all}
	\end{subfigure}
	\begin{subfigure}{\textwidth}
		\centering
		\includegraphics[width=0.8\textwidth]{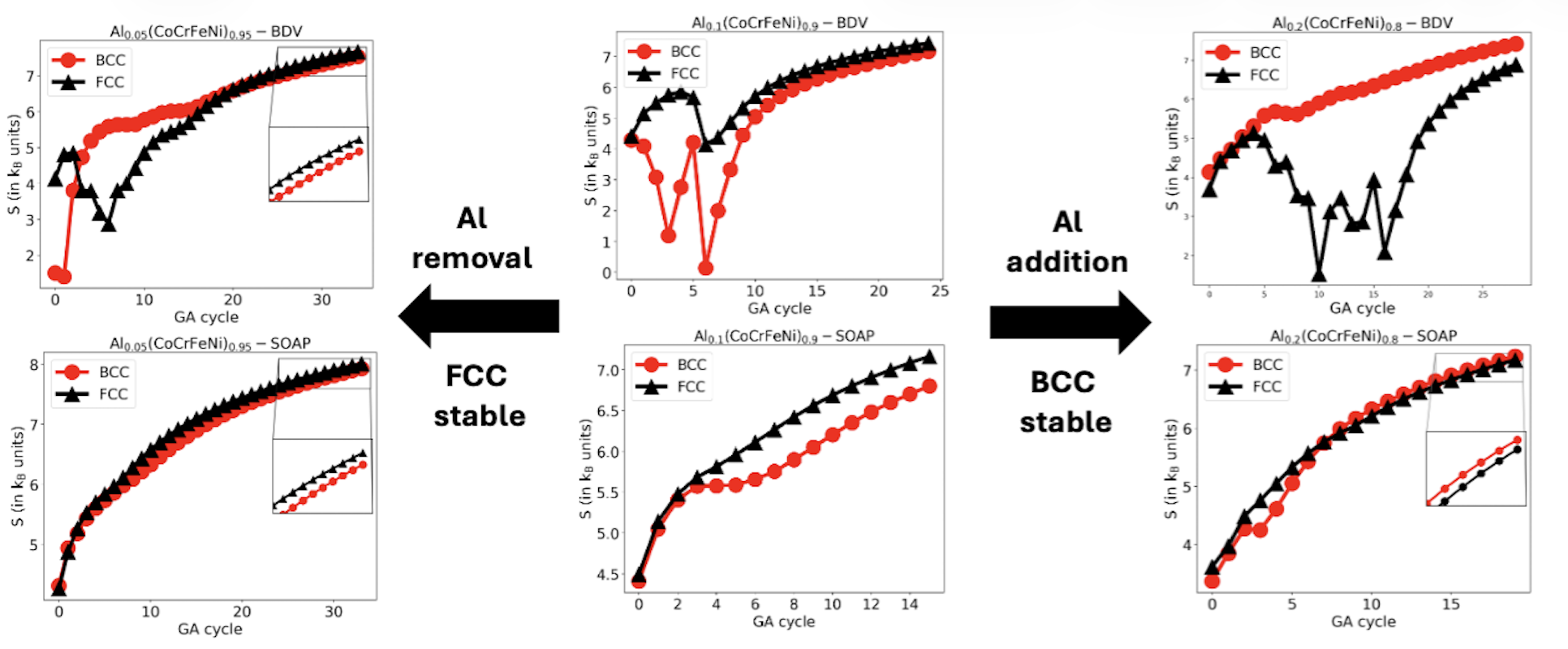}
		\caption{}
		\label{subfig:soapbdv-pt}
	\end{subfigure}
	\caption{(a) The variation of the entropy parameter for the candidate crystal structures (BCC and FCC) for binary, ternary, and quaternary alloys and (b) the application of the entropy parameter for the phase prediction with composition variation of $\mathrm{Al_{x}(CoCrFeNi)_{1-x}}$}.
	\label{fig:soap-bdv}
\end{figure*}

The concept of information entropy was originally developed by Claude Shannon \cite{shannon1948mathematical} in the context of information theory. The quintessential noise in the information exchange exhibited a conceptual analogy to thermodynamic entropy, which encompasses energy dissipation in the thermodynamic state space. Information entropy has been extensively applied not only in information theory but also across a range of disciplines, including ecology \cite{pos2023unraveling}, health sciences \cite{samadian2025entropy}, and quantum systems \cite{yatomi2025quantum}, among others. \\
The rigorous studies on the relationship between thermodynamic entropy and information entropy \cite{parrondo2015thermodynamics} paved the way for the application of information entropy in materials systems. The information entropy concept has been employed for studying the complexity of molecules by Krivovichev \cite{krivovichev2012topological,krivovichev2016structural}. A correlation between the information entropy and the thermodynamic entropy was discussed. Such an approach has been further extended for molecular systems \cite{hornfeck2020extension,kaussler2021crystit}. This approach is an interesting method for determining the complexity of a molecule, and such studies pave the way for employing information entropy parameters to study physical phenomena. Information entropy has been employed as a descriptor to study the nucleation of molecular crystals \cite{gobbo2018nucleation} and molecular cocrystals \cite{song2020generating}. The concept of information entropy was applied to the Sudoku-inspired lattice occupancy of high-entropy alloys \cite{zhuang2022sudoku}. \\
Figure \ref{subfig:pdfall} shows the schematic of the reference energy distribution (which corresponds to the thermodynamic systems) and the sampled potential energy distribution. The difference between the reference $\mathrm{(Eq)}$ and the sampled energy distribution $\mathrm{(P)}$ can be mathematically expressed as the Kullback-Leibler (KL) divergence ($\mathrm{D_{KL}}$),  which quantifies the information loss between the distribution, expressed as,
\begin{equation}
\mathrm{
	D_{KL} \left(P_x \parallel Eq_x\right)= -\sum_{x \in \chi} P_x \cdot log \left(\frac{Eq_x}{P_x}\right)
	}
\end{equation}
where, $\mathrm{P_x}$ and $\mathrm{Eq_x}$ are the sampled and the referece distributions, respectively. Though the above expression is defined for arbitrary distributions, in our case, the reference distribution is the equilibrium distribution. Note that the $\mathrm{\chi}$ is the sample space and the summation sign is due to the discrete nature of the samples.  Figure \ref{subfig:kld} shows the schematic of the $\mathrm{D_{KL}}$ graphically. It is clear that the higher value of the $\mathrm{D_{KL}}$ corresponds to the higher information loss and \emph{vice-versa}. \\
In the problem involving the $\mathrm{D_{KL}}$ for the range of distributions with respect to the reference distribution, the mathematical task is to compare the distributions. Since the sampled energy distributions are discrete in nature, similarity between the candidate distribution and the equilibrium distribution may be, in principle, determined using the widely used Kernel methods with the \emph{kernel trick}. Kernel methods can help find similarities between distributions by implicitly operating in a higher-dimensional feature space via kernel functions. In this scenario, the similarity between the sampled and the equilibrium distribution needs to be calculated. However, this would require sampling the equilibrium distribution, which implies knowledge of the equilibrium crystal structure (which is not known beforehand). The sampling of the equilibrium distribution can be carried out using \emph{ab initio} methods exhibiting chemical accuracy (\emph{i.e.,} $\mathrm{10^{-3}\: eV/atom}$) with the Metropolis Monte Carlo method, in principle. But this method is prohibitively computationally expensive, rendering it unsuitable for efficient phase stability studies of chemically disordered systems in general.\\
In view of the above, we aim to employ the $\mathrm{D_{KL}}$ for the phase prediction using the comparative $\mathrm{D_{KL}}$  for candidate crystal structures (P and Q as shown in Fig. \ref{subfig:unbound}). The $\mathrm{D_{KL}}$ is chosen as a metric due to the high computational cost of thermodynamic entropy and free energy calculations. It is understood that $\mathrm{D_{KL}}$  is inversely related to the thermodynamic entropy. In this work, we propose predicting the phase from candidate crystal structures by biased sampling of thermodynamically relevant low-energy states, treating this as an information alignment problem. The $\mathrm{D_{KL}}$ difference between the candidate crystal structure, P and Q, may be represented as (see appendix for the derivation),
\begin{equation}
	\mathrm{
	D_{KL}(P || Eq) - D_{KL}(Q || Eq) = H(Q) - H(P)
	}
\end{equation}
where H(Q) and H(P) are the Shannon entropies of the candidate crystal structures, Q and P, respectively. The H is simply $\mathrm{-\sum_{i} {\left(p_i \cdot \ln(p_i) \right)}}$. The $\mathrm{p_i}$ or the probability of the existence of a microstate is expressed as,
\begin{equation}
	\mathrm{
		p_i = \frac{ \exp(-(E_i - E_{min})) }{\sum_{i} { \exp(-(E_i - E_{min}))  }}
	}
\end{equation}
where $\mathrm{E_i}$ and $\mathrm{E_{min}}$  are the potential energy of the microstate as calculated in the canonical ensemble and minimum energy as determined from the atomistic sampling, respectively. Note that $\mathrm{Eq}$ distribution in the present context is the Boltzmann distribution, which maximises the entropy (and hence minimises the $\mathrm{T\cdot S}$ term of the free energy function) and minimises the internal energy, thus minimising the free energy function as a whole, if the expectation value of the internal energy is constant and $\mathrm{\sum_{i}{ p_i }} = 1$ \cite{gao2019generalized}.\\
In the present work, the candidate crystal structures can exhibit their own energy bound during the biased alchemical Monte Carlo using the GAASP approach \cite{anand2023gaasp}. The candidate crystal structure with a minimum $\mathrm{D_{KL}}$ with the Boltzmann distribution should correspond to the equilibrium crystal structure. However, this cannot be assured with mathematical accuracy. In view of this, we propose to employ the Shannon Entropy difference as a metric of the phase prediction of chemically disordered alloys. Figure \ref{subfig:soapbdv-all} shows the variation in the entropy parameter for candidate crystal structures (BCC and FCC) binary (CoNi, MoW, FeNi and TaW), ternary (CoCrNi and CrFeNi), quaternary (CoCrFeNi) alloys (Fig. \ref{subfig:soapbdv-all}). It is clear that the entropy parameter at 300K can predict phase stability for the range of alloys at the temperature of interest, as at 300K, FCC phase is shown to have higher Shannon entropy for CoNi, FeNi, CoCrNi, CrFeNi, and CoCrFeNi, while the BCC phase has higher Shannon entropy for MoW and TaW alloys, The GCNN model with both BDV and SOAP descriptors shows similar qualitative performance in phase prediction. We further explored the effect of the composition on the phase stability of $\mathrm{Al_{x}(CoCrFeNi)_{1-x}}$ with $\mathrm{x = 0.05,\: 0.1,\: and \: 0.2}$, as shown in the Fig. \ref{subfig:soapbdv-pt}. Our predictions are in line with the published reports \cite{liu2026situ,tabassum2025effect,panda2021studies}.

\FloatBarrier

\section{Conclusions}
The present investigation demonstrates the applicability of the Alchemical Monte Carlo (or GAASP) computational framework, combined with a Graph Convolutional Neural Network, for atomistic sampling, efficient potential-energy calculation, and exploration of the thermodynamically relevant potential-energy landscape (PEL), followed by Information or Shannon entropy calculations on the explored PEL. The following conclusions can be drawn:
\begin{itemize}
	\item  The atomistic descriptor design for efficient machine learning (ML) is dependent on the complexity of the material being studied. The simpler Bond Disproportion Vector (BDV) shows the comparative performance in comparison with the Smooth Overlap of Atomic Positions (SOAP) descriptor for chemically disordered ternary, quaternary, and quinary alloys, while the SOAP descriptor performs well for the binary chemically disordered alloys for predicting the per-atom potential energy as a function of its coordination environment.
	\item	 Comparatively simple Graph Convolution Neural Network (GCNN) with two hidden layers, shows the predictive capability of a complex potential energy landscape of chemically disordered alloys with explicit atomic descriptors (SOAP and BDV) and Term Frequency-Inverse Document Frequency (TF-IDF) featurisation,
	\item The Information entropy defined on the PEL as deduced from the evolutionary Monte Carlo with GCNN ML model can be employed for the efficient phase prediction of chemically disordered alloys, including high-entropy alloys. 
\end{itemize}

\section{Acknowledgments}
The support and the resources provided by PARAM Shivay Facility under the National Supercomputing Mission, Government of India, at the Indian Institute of Technology, Varanasi, are gratefully acknowledged. GA is thankful to Prof P. D. Gujrati for the helpful discussions.

\section{Code Availability}
The GAASP code for the alchemical Monte Carlo sampling is available at https://github.com/ganand1990/GAASP, and the scripts for the Graph Convolutional Network are available at https://github.com/ganand1990/GCNN.
\bibliographystyle{unsrt}
\bibliography{manuscript}

\appendix
\renewcommand{\theequation}{A.\arabic{equation}}
\setcounter{equation}{0}
\section*{Appendix}
\subsection*{Relation between comparative Shannon Entropy and Kullback-Leibler Divergence (KLD)}
The Shannon Entropy can be defined as,
\begin{equation}
	\mathrm{
	H(P) = -\sum_x{P(x) \cdot	\ln{P(x)}}
	}
\end{equation}
where $\mathrm{P(x)}$ is the discrete distribution and the KL divergence with respect to the reference distribution $\mathrm{Eq(x)}$ is defined as,
\begin{equation}
	\mathrm{
		D_{KL}(P || Eq) = -\sum{P(x) \ln \left({\frac{Eq(x)}{P(x)}}\right)}
	}
\end{equation}	
Above may be rearranged as,
\begin{equation}
	\mathrm{
		D_{KL}(P || Eq)  = H(P,Eq) - H(P)
	}
\end{equation}		
where, $\mathrm{H(P,Eq)}$ is equal to $\mathrm{-\sum{P(x) \cdot \ln \left(Eq(x) \right)}}$ and this terms is known as \emph{cross-entropy}. We may be able to write the KL divergence for another distribution, $\mathrm{Q}$, as,
\begin{equation}
	\mathrm{
		D_{KL}(Q || Eq)  = H(Q,Eq) - H(Q)
	}
\end{equation}	
The difference between the KL divergences of P and Q may be expressed as,
\begin{equation}
	\mathrm{
		D_{KL}(P || Eq)  - D_{KL}(Q || Eq) = H(Q) - H(P)
	}
\end{equation}	

\end{document}